\documentclass[Journal,a4paper]{ascelike-new}
\usepackage[utf8]{inputenc}
\usepackage[T1]{fontenc}
\usepackage{lmodern}
\usepackage[figurename=Fig.,labelfont=bf,labelsep=period]{caption}
\usepackage{graphicx}
\usepackage{lscape}
\usepackage{relsize}
\usepackage{rotating}
\usepackage{color}
\usepackage{amsfonts,amsmath,amssymb,bm}
\usepackage[compress]{natbib}
\usepackage{xcolor}
\usepackage{multirow}
\usepackage{threeparttable,subcaption}
\usepackage{newtxtext,newtxmath}
\usepackage{algorithm,tikz}
\usepackage[]{algpseudocode}
\usepackage{pgfplots}
\usepackage[colorlinks=true,citecolor=red,linkcolor=black]{hyperref}
\usepackage{nomencl} 
\makenomenclature
\usepackage{etoolbox}
\usepackage{array}
\usepackage{subcaption,graphicx}
\usepackage{nicematrix}
\usepackage{multirow}

\begin{document}
\nolinenumbers

\title{Accelerated Development of Multi-component Alloys in Discrete Design Space Using Bayesian Multi-Objective Optimization}

\author[1]{Osman Mamun}
\author[2]{Markus Bause}
\author[1,2,*]{Bhuiyan Shameem Mahmood Ebna Hai}

\affil[1]{Fehrmann MaterialsX GmbH - Fehrmann Tech Group, Hamburg, Germany}
\affil[2]{Helmut Schmidt University - University of the Federal Armed Forces Hamburg, Germany}
\affil[*]{Corresponding author: shameem.ebna.hai@fehrmann-materialsx.com}


\maketitle

\begin{abstract}
    Bayesian optimization (BO) protocol based on Active Learning (AL) principles has
    garnered significant attention due to its ability to optimize black-box objective
    functions efficiently. This capability is a prerequisite for guiding autonomous
    and high-throughput materials design and discovery processes. However, its application
    in materials science, particularly for novel alloy designs with multiple targeted
    properties, remains limited. This limitation is due to the computational complexity
    and the lack of reliable and robust acquisition functions for multiobjective
    optimization. In recent years, expected hypervolume-based geometrical acquisition
    functions have demonstrated superior performance and speed compared to other
    multiobjective optimization algorithms, such as Thompson Sampling Efficient
    Multiobjective Optimization (TSEMO), Pareto Efficient Global Optimization (parEGO), etc.
    This work compares several state-of-the-art multiobjective BO acquisition functions,
    i.e., parallel expected hypervolume improvement (qEHVI), noisy parallel expected
    hypervolume improvement (qNEHVI), parallel Pareto efficient global optimization (parEGO),
    and parallel noisy Pareto efficient global optimization (qNparEGO) for the
    multiobjective optimization of physical properties in multi-component alloys.
    We demonstrate the impressive performance of the qEHVI acquisition function
    in finding the optimum Pareto front in 1-, 2-, and 3-objective Aluminium
    alloy optimization problems within a limited evaluation budget and reasonable
    computational cost. In addition, we discuss the role of different surrogate
    model optimization methods from a computational cost and efficiency perspective.
    Finally, we illustrate the efficiency of the pool-based active learning protocol
    to further accelerate the discovery process by conducting several computational
    and experimental campaigns in each iteration. This approach is particularly
    advantageous for use in massively parallel high-throughput synthesis facilities
    and computer architectures.
\end{abstract}

\section{Introduction}

In materials science, when designing suitable materials for a
specific application from a broad range of alternatives, a researcher must
consider factors such as individual characteristics, applications, advantages,
and limitations. A lack of understanding of the functional requirements and the
specific design criteria that must be met can lead to the development of
sub-optimal materials. In turn, the lack of proper optimization can result in
significant costs or wasted investment, leading to premature component or product
failure. For example, gas turbine engines must operate at high temperatures
withstanding high pressures to achieve high efficiency. This extreme condition
not only improves the thermal efficiency of the process but also benefits the
environment by helping to reduce carbon emissions \cite{long2018microstructural}.
Austenitic stainless steel and Ni-based superalloys are widely used in gas turbines
due to their excellent corrosion resistance and high yield and tensile strength
at elevated temperatures and pressures \cite{mamun2021machine}. However, robust
corrosion resistance and high strength are not the only criteria to be met; an
ideal material for gas turbine engines should also have low density, high thermal
conductivity, and a small coefficient of linear thermal expansion \cite{khatamsaz2023bayesian}.
Likewise, in the design of a multi-principal-element alloy (MPEA), several
objectives must be fulfilled, including low-density, high-temperature yield
strength, creep resistance, and oxidation resistance \cite{liu2023machine}.

Designing complex materials with desired properties requires expensive
computational modeling, such as density functional theory (DFT) computations,
thermochemical simulations, molecular dynamics calculations, and
extensive experimental campaigns. The highly non-linear interactions between
design variables and target properties mean that high-fidelity models often
behave like black-box functions. This behavior hinders the ability to design
tailored materials, which usually necessitates carrying out these costly computations
or experiments for a wide range of candidate materials to optimize several
competing properties. A combinatorial approach to such problems involves conducting
simulations or experiments on an exhaustive combination of numerous variables,
including compositions, processing, and synthesis parameters. This approach aims
to establish a structure-property relationship that can guide scientists in
understanding the interplay of different parameters \cite{GEBHARDT20125491},
potentially leading to the development of state-of-the-art novel materials.
However, this method needs more scalability and is time-consuming \cite{e16094749}.
Estimates suggest it can take a few decades to productionize
novel materials from the ideation phase to the market \cite{agrawal2016perspective}.
Autonomous experimental systems have emerged as powerful tools to
expedite the lab-to-market process. These systems function as oracles, suggesting
the following experiment to be performed sequentially, thereby avoiding the
combinatorial approach and saving time and resources.

Autonomous experimental and computational systems are becoming the frontier of
accelerated materials design \cite{liang2021benchmarking}. In this context,
active learning has emerged as a revolutionary tool, significantly contributing
to discovering novel materials while minimizing experimental and computational
costs. The primary advantage of active learning is its ability to iteratively
guide the selection of experiments toward a target design objective, thus
reducing the number of required experiments. Bayesian optimization, a class of
active learning methods, employs a surrogate model and a utility function to
recommend the subsequent experiment or computation \cite{snoek2012practical}.
This surrogate model correlates design variables, such as materials composition,
with target properties. The utility function then quantifies the impact of various
unknown parameters on these target properties, guiding the selection of the following
experiment(s) to maximize the utility function's value. Bayesian optimization
offers several advantages in experimental design, including sample efficiency,
global optimization, and the ability to handle complex objective functions without
a closed form \cite{7352306}. As a result, Bayesian optimization (BO) has become
an increasingly attractive choice for accelerating materials research and optimizing
material properties beyond current state-of-the-art technologies, as evidenced by
the plethora of publications involving the application of BO in materials science.

Although Bayesian optimization (BO) has demonstrated promising outcomes in the
materials science field, its implementation is still restricted due to several factors:
\begin{enumerate}
    \item[(a)] It is primarily employed for single-objective optimization.
    \item[(b)] Only a single experiment or computational recommendation is feasible
          in each iteration, which constrains the application of parallel computation or
          experimentation strategies for improved data collection.
    \item[(c)] Surrogate probabilistic modeling often proves expensive and
          impractical in contemporary computer architectures. For instance, the Gaussian
          Process, a common choice for surrogate modeling, scales as \(N^3\), where \(N\)
          represents the number of data instances.
\end{enumerate}

Single-objective optimization can be achieved using acquisition functions such
as the Upper Confidence Bound (UCB), Lower Confidence Bound (LCB), Expected
Improvement (EI), and Probability of Improvement (PI) \cite{frazier2018tutorial}.
However, multiobjective optimization, which involves simultaneously optimizing
several competing properties, adding a layer of complexity. To address this,
several approaches have been proposed. For instance, Thompson
Sampling for Efficient Multiobjective Optimisation (TSEMO), a multiobjective
evolutionary algorithm \cite{bradford2018efficient}; Pareto Efficient Global
Optimization (parEGO), extending the expected improvement acquisition function
to multi-dimensional problems \cite{knowles2006parego}; and Expected Hypervolume
Improvement (EHVI), a hypervolume maximization algorithm \cite{hupkens2015faster}.
Among these, EHVI has gained significant attention due to its simplicity and
effectiveness. This method aims to maximize the hypervolume covered by the
non-dominated points relative to a fixed reference point, effectively identifying
a well-defined Pareto front.

However, like TSEMO and parEGO, EHVI is notoriously costly and does not
support parallel computations. The parallel expected hypervolume improvement (qEHVI)
and parallel noisy expected hypervolume improvement methods overcome these
limitations by enabling the parallel calculation of the hypervolume for different
points using the inclusion-exclusion principle \cite{daulton2020differentiable}.
Similarly, the parallel parEGO and its noisy variant parallelize the scalarization
process, thereby enabling Bayesian optimization for multi-dimensional problems.
Having access to these acquisition functions for multiobjective optimization,
the question naturally arises of how to select the best algorithm for the application
of novel materials design and discovery. Specifically, there are two important
factors to consider: 1. the surrogate model construction, and 2. the acquisition
function selection.

Another challenge in Bayesian Optimization (BO) within materials science is the
common practice of making only a single recommendation per iteration, which is
inefficient when multiple parallel experiments or computations are possible.
To address this, several strategies have been suggested, such as dividing the
objective space into clusters based on similarity or distance vectors to maximize
data diversity and informativeness \cite{mamun2022uncertainty, zhan2021comparative}.
The qEHVI also supports the generation of multiple candidate points simultaneously,
ensuring proper propagation of uncertainty.

This work presents a Bayesian optimization (BO) framework to efficiently
optimize multiple properties simultaneously for Aluminium alloy design problems.
First, we compare the performance of two ways of constructing or
conditioning the surrogate model on the training data: maximum likelihood
estimation and Monte Carlo simulation. Our analysis shows that
maximum likelihood estimation provides the best trade-off for accuracy and time
constraints. Then, we compare three different acquisition functions for 1-dimensional
properties optimization, i.e., probability of improvement (PI),
expected improvement (EI), and expected hypervolume improvement (EHVI).
Our results illustrate that all three acquisition functions perform more or
less similarly and vastly outperform the random data acquisition algorithm, with
PI being slightly worse than the other two algorithms.
Next, we investigate the performance of four different acquisition functions for
multi-dimensional optimization problems, i.e., expected hypervolume
improvement (EHVI), noisy expected hypervolume improvement (NEHVI),
Pareto efficient global optimization (parEGO) and noisy Pareto efficient global
optimization (NparEGO). Except for parEGO, the remaining three algorithms outclasses
the random data acquisition algorithm. However, when considering the computational
cost of computing the utility value with the noisy variant of these algorithms,
qEHVI becomes an automatic choice for multi-dimensional problems; however, for
noisy data (e.g., experimental observations), qNEHVI should be preferred to deal
with the observation uncertainty. We also demonstrate that the EHVI can identify the
global Pareto front with only a 70-75 evaluation budget and achieves a hypervolume
that is over 90\% of the maximum achievable value. Finally, we demonstrate that
multiple data acquisition in each iteration cycle can lead to enhanced performance,
in terms of time to reach a predefined hypervolume, ultimately leading to faster
alloy design and discovery.

\section{Methods}

In this work, we present an active learning framework for the multi-objective optimization
of Aluminium alloy design and discovery. Below, we briefly describe the underlying theory
of the active learning framework, including surrogate model construction and various
single- and multi-objective acquisition functions.

\begin{figure}
    \captionsetup{justification=centering}
    \includegraphics[clip, width=\columnwidth]{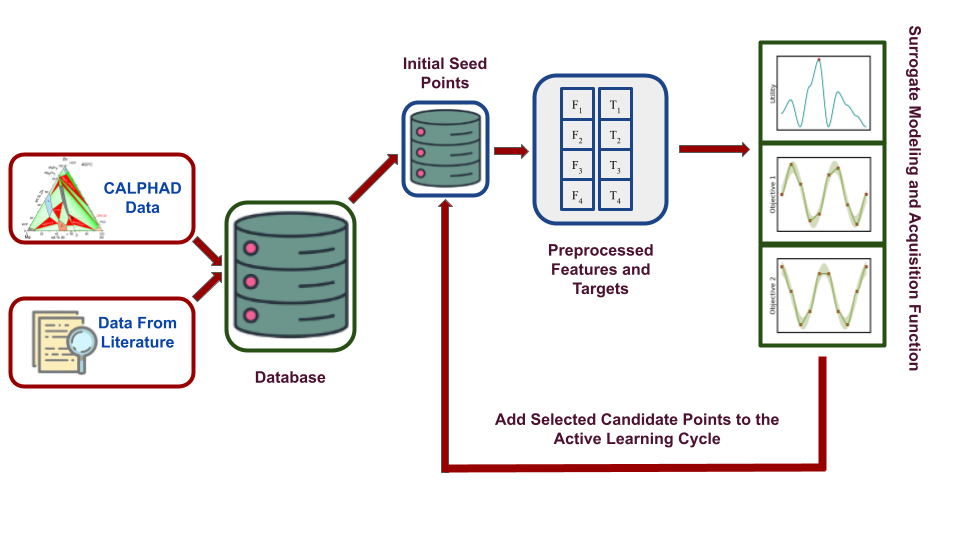}
    \caption{The workflow to train multi-objective Bayesian active learning
        framework for alloy properties optimization}\label{fig:workflow}
\end{figure}

Figure \ref{fig:workflow} illustrates the general workflow of the active learning
cycle. The cycle begins with observed data for 'n' cases, featuring variables
\{$F_1, F_2, ..., F_n$\} and targets \{$T_1, T_2, ..., T_n$\}. To simulate a
realistic situation, we start with a fixed number of initial points (e.g., 25, 50, etc.)
acquired at random. The initial seed points then undergo a data preparation stage where:
\begin{enumerate}
    \item[(a)] For experimental data, data instances where elongation is greater
          than 60\% are removed as they might contain experimental measurement errors.
          Since our objective is to minimize the elongation in most of the realistic
          scenarios, this data processing will have no significant impact on the model bias.
    \item[(b)] Any feature with null values is removed, and
    \item[(c)] The data is scaled between zero and one to remove numerical scale bias.
\end{enumerate}

A probabilistic model is subsequently trained on these data points, enabling the
inference of data instances where only the feature vector is known. In multi-objective
optimization, either a single model can be used to model the joint probability
distribution of the target properties or separate models can be trained for each
target. Based on the model's inferences for both single- and multi-objective problems,
the utility value is computed using acquisition functions for the discrete data instances
for which the features are known but the target properties are not. The algorithm
then selects the point or points (in cases where several candidates are chosen in
each iteration) that maximize the utility function. These recommended points are
then added to the existing training database. The cycle continues until it reaches
the optimum point. For single-objective optimization, the minimum or maximum value is
considered the optimum and used as the stopping criterion. For multi-objective
optimization, the stopping criterion is the hypervolume covered by the
non-dominated points in the entire database. In situations where the global
maximum hypervolume is not known a priori, the cycle can continue until it
reaches a predetermined number of iterations or a specified hypervolume that
meets specific design criteria.

Based on the above discussion, we can identify two key components in an active learning cycle:
\begin{enumerate}
    \item[(a)] A probabilistic model, serving as a surrogate model, to
          approximate the underlying data-generating function. In this article,
          we use Gaussian processes, which are particularly effective in
          multi-objective optimization.
    \item[(b)] An acquisition function is used to quantify the utility of all the
          candidate data points, which then guides the selection of the next data
          point to evaluate.
\end{enumerate}

\subsection*{Gaussian Process}

The Gaussian process (GP), a powerful non-parametric machine learning algorithm,
has become the workhorse in Bayesian optimization due to its flexibility and
excellent uncertainty quantification capability. To estimate the mean \(\hat{\mu} (x^*)\)
and standard deviation \(\hat{\sigma} (x^*)\) of an unobserved data point \(x^*\),
the Gaussian process is employed. It defines the joint distribution of the observed
and unobserved objective values as follows:

\begin{equation}
    \left[\begin{array}{c}
            \mathbf{y} \\
            y_*
        \end{array}\right] \sim \mathcal{N}\left(0, \left[\begin{array}{cc}
            K + \sigma^2 I & K_*^T  \\
            K_*            & K_{**}
        \end{array}\right]\right)
\end{equation}

where \( \mathbf{y} \) represents the observed values, \( y_* \) the unobserved
values, and \( K \), \( K_* \), and \( K_{**} \) are components of the kernel
matrix. Specifically, \(K + \sigma^2 I\) is the covariance matrix of the observed
observations, \(K_*^T\) and \(K_*\) represent the covariances between observed
and unobserved observations, and \(K_{**}\) is the covariance matrix of the
unobserved observations.

In this article, we utilise the Matérn 5/2 kernel function, a widely employed
kernel function renowned for its useful mathematical properties. The Matérn 5/2
kernel for two points \( p \) and \( q \) in the design space is defined as:

\begin{equation}
    k(\mathbf{p}_j, \mathbf{q}_j) = \sigma_0^2 \cdot \left(1 + \frac{\sqrt{5}r}{l_j} + \frac{5r^2}{3l_j^2}\right) \exp\left(-\frac{\sqrt{5}r}{l_j}\right)
\end{equation}

where \( r = \sqrt{(p_j - q_j)^2} \) and \( l \) is the characteristic length.

The mean and variance of the posterior are computed as follows:

\begin{equation}
    \hat{\mu}(\mathbf{x}) = y_* = K_* [K + \sigma^2 I]^{-1} \mathbf{y}
\end{equation}

and

\begin{equation}
    \text{cov}(y_*) = K_{**} - K_* [K + \sigma^2 I]^{-1} K_*^T
\end{equation}

To make inferences on unknown design features, we utilize two methods:
\begin{enumerate}
    \item[(a)] Maximum a Posteriori (MAP), which formulates the GP problem as an
          optimization problem with respect to the target density, resulting in faster
          model training and inference compared to No U-Turn Samplers (NUTS).
    \item[(b)] No U-Turn Samplers (NUTS), a full Monte-Carlo method that updates
          the posterior via an acceptance-rejection algorithm, making model training
          and inference more resource-intensive.
\end{enumerate}

For MAP training, we utilized two different open source implementations:
\begin{enumerate}
    \item[(a)] scikit-learn, more commonly known as sklearn. We used the sklearn's
          \textit{GaussianProcessRegressor} API to train the model. Since sklearn's model API
          is not compatible with the BoTorch acquisition function API (which we used to
          implement all the acquisition functions in this study), we provide a custom
          wrapper to make the sklearn model compatible with the BoTorch acquisition
          function API.
    \item[(b)] BoTorch's native \textit{SingleTaskGP} model API along with
          \textit{ModelList} API for multiobjective model training.
\end{enumerate}

Apart from using the aforementioned two models, we also used sparse axis-aligned
subspace GP (SAASGP), which, unlike predefined inflexible kernels, utilizes a flexible
kernel definition to systematically reduce the number of parameters to train while
capturing the anisotropic nature of the feature space.

For a large number of dimensions, Gaussian Processes (GP) scale as \( N^3 \) and
become computationally very expensive, particularly when assuming a certain
degree of smoothness. To enhance the sample efficiency of GP, several assumptions
are made in the Sparse Axis-Aligned Subspaces (SAAS) GP formulation \cite{eriksson2021high}:
\begin{enumerate}
    \item[(a)] Assume a hierarchy of feature relevance to reduce the number of active features systematically.
    \item[(b)] Incorporate a flexible class of smooth, non-linear functions.
    \item[(c)] Ensure tractable inference due to the differentiable nature of the kernel functions.
\end{enumerate}

To facilitate the assumptions mentioned above in Sparse Axis Aligned Subspaces (SAAS),
the Gaussian Process (GP) models are initialized with a structured prior over
the kernel hyperparameters. These include:
\begin{enumerate}
    \item[(a)] A weak prior over the kernel variance.
    \item[(b)] A half-Cauchy prior on the global shrinkage or regularization parameter, leading to the zeroing out of irrelevant features.
    \item[(c)] A half-Cauchy prior on the (inverse squared) length scale for Radial Basis Function (RBF) or Matern kernels.
\end{enumerate}

Due to these modifications, most dimensions are effectively switched off (their
contributions become negligible or zero) in line with Automatic Relevance Determination
(ARD). The half-Cauchy prior over the length scale causes these values to cluster
around zero unless the data provides sufficient evidence to increase them. By design,
the model remains parsimonious; however, as more data is gathered, some model
parameters may deviate from zero, resulting in an excellently regularized model.
The SAASGP models are trained with Monte-Carlo integration along with No U-turn
samplers (NUTS).

\subsection*{Acquistion Functions}

\subsubsection*{Probability of Improvement (PI)}

If we denote the $f^*$ to be the maximum obtained so far in the context of a maximization
problem, while $\Phi(.)$ and $\phi(.)$ represent the normal cumulative distribution
function (CDF) and the probability density function (PDF), respectively, of the
standard normal random variable. Then, the probability of improvement over the current
best point is defined as:

\begin{equation}
    PI(x) = P(f(x) \ge f(x^*)) = \Phi\left(\frac{\mu(x) - f(x^*)}{\sigma(x)}\right)
\end{equation}

Here, PI is the probability of finding a better function value at position $ x $
than the current best value ($ f(x^*) $). In simple terms, whenever a point $ x $
offers an improvement over the current best point; it is given a reward (or utility
function value); otherwise no reward at all. Then, the probability of improvement
is computed as the expected utility. In this work, we use the parallel implementation
of the PI acquisition function, which will be dubbed qPI in the rest of the article.

\subsubsection*{Expected Improvement (EI)}

The probability of improvement acquisition function has a serious flaw in that
it only considers the function value at position $x$ that is better than
the current best point, irrespective of the magnitude of the improvement, which leads
to suboptimal performance as the algorithm often gets trapped in a local optima.
To overcome this limitation, expected improvement (EI) was proposed, which not only takes
the improvement over the current best into account, also the magnitude of the improvement
is taken into account to distinguish the global optimum from the local optimums. EI is
computed as:

\begin{equation}
    EI(x) = \mathbb{E}[max(0, f^* - f(x))] = (f^* - \mu(x)) \Phi\left(\frac{f^* - \mu(x)}{\sigma(x)}\right) + \sigma(x) \phi\left(\frac{f^* - \mu(x)}{\sigma(x)}\right)
\end{equation}

where $\mathbb{E}$ denotes the expectation value, $\Phi$ and $\phi$ are the Gaussian
CDF and PDF, respectively. In this article, we used the parallel implementation of the
EI acquisition function, which shall be referred to as qEI in the subsequent analysis.

\subsubsection*{Expected Hypervolume Improvement (EHVI)}

For a given Pareto front, denoted as \(\mathcal{P}\), the hypervolume (HV) is
defined as the d-dimensional Lebesgue measure of the corresponding subspace.
In our multi-objective optimization approach, we utilized the expected hypervolume
improvement as the acquisition function for the active learning cycle. For a
Pareto set \(P\) and a reference point \(r\), the hypervolume improvement of a
vector \(y\) is defined as:

\begin{equation}
    HVI(\mathcal{P}, y) = HV(\mathcal{P} \cup \{y\}) - HV(\mathcal{P}).
\end{equation}

The expected hypervolume is the expectation of the HVI over the posterior
distribution, which can be estimated using the Monte Carlo integration.

\begin{equation}
    \alpha_{EHVI (x | \mathcal{P})} = \mathbb{E}[HVI(f(x) | \mathcal{P})]
\end{equation}

In this work, we used The q-expected hypervolume improvement (qEHVI), a parallel
implementation of the EHVI algorithm developed by Facebook \cite{daulton2020differentiable},
which introduces several modifications for practical implementation. Unlike the original
EHVI, which utilizes gradient-free acquisition function optimization, qEHVI leverages
auto-differentiation to facilitate the optimization of the acquisition function
using first-order and quasi-second-order methods. Additionally, it enables the
computation of EHVI on modern parallel computing architectures, significantly
increasing speed and efficiency. Owing to these significant improvements, qEHVI
outperformed other multi-objective Bayesian optimization algorithms in many
standard benchmark problems.

\subsubsection*{Noisy Expected Hypervolume Improvement (NEHVI)}

In the EHVI or qEHVI algorithm, one of the major assumptions is that the observations
are noise-free; unfortunately, in many real-world applications, this assumption
seriously challenges the performance of the qEHVI for efficient Bayesian optimization.
Instead of using only the posterior mean to compute the expected hypervolume improvement
of different candidate points, the posterior uncertainty is also taken into account to
compute the acquisition function values. The acquisition function, noisy expected hypervolume
improvement (NEHVI) is defined as:

\begin{equation}
    \alpha_{NEHVI(x)} = \int \alpha_{EHVI}(x|\mathcal{P}_{n})p(f|D_n)df
\end{equation}
where $P_n$ denotes the Pareto frontier over $f(X_n)$. Using MC integration, the
inner expectation of $\alpha_{NEHVI(x)}$ is computed. The parallel version of
this algorithm is dubbed as qNEHVI \cite{daulton2021parallel}.

\subsubsection*{Pareto Efficient Global Optimization (parEGO)}

Pareto Efficient Global Optimization (parEGO) is a multiobjective optimization
algorithm where the multiobjective optimization problem is transformed into a single
objective optimization problem using a suitable scalarization algorithm. Then it
applies the expected improvement (EI) acquisition function (described above) to
compute the utility of candidate points. For parEGO algorithm, the random augmented
Chebyshev scalarizations technique is used \cite{knowles2006parego}. BoTorch implemented a
novel parallel version of this algorithm, qparEGO, which is used in this article.

\subsubsection*{Noisy Pareto Efficient Global Optimization (NparEGO)}

Similar to the qNEHVI algorithm discussed above, the noisy variant of the parEGO
(dubbed as NparEGO) utilizes the posterior mean and standard deviation to
compute the true Pareto frontier. In this article, we also use the parallel version
of this algorithm, known as qNparEGO.

\subsection*{Multiple Data Acquisition}

There are several ways to acquire multiple data in each iteration. In this article,
we use the sequential greedy batch selection algorithm as implemented in the
BoTorch code. In a sequential greedy batch selection algorithm, the utility values
of all the candidate points are evaluated first, and then the best candidate point
is selected. Then, for subsequent candidate selection, the previously selected candidates
are held constant to compute the utility values of the remaining candidate points.

\section{Results and Discussion}

\subsection*{Dataset}
In this study, two different datasets were utilized:
\begin{enumerate}
    \item[(a)] Computational Dataset: This dataset was exclusively generated using
          CALPHAD simulation software for this study. It comprises 13,700 data instances
          with 41 features (after preprocessing steps). Detailed information regarding the
          simulation protocol and dataset properties are provided in Appendix A of
          the Supporting Information.
    \item[(b)] Experimental Dataset: This data was gathered and compiled into a
          consistent and reliable dataset by peer-reviewed literature \cite{Pfeiffer2022}
          and Fehrmann MaterialsX GmbH. It includes data from experiments conducted
          in-house and other credible sources, such as published articles and a European
          consortium. The dataset comprises approximately 3,700 data instances with 13
          features (after preprocessing steps), collected over several years. These
          features cover the composition of elements and processing parameters like
          temper and temperature. Detailed information on the dataset properties is
          illustrated in Appendix B of the Supporting Information.
\end{enumerate}

\subsection*{Surrogate Model Construction}

For surrogate model training, we used BoTorch's\cite{balandat2020botorch} \textit{SingleTaskGP}
model API (BT-MLE) and scikit learn's\cite{scikit-learn} \textit{GaussianProcessRegressor}
API (SK-MLE). As for the kernel, we used a combination of Matern kernel along with
a linear kernel. For optimization of both the models, we used scipy's\cite{2020SciPy-NMeth}
optimizer module. For SAASGP model training, we used BoTorch's \textit{SaasFullyBayesianSingleTaskGP}
model API. To train the model, we used the GPyTorch's\cite{gardner2018gpytorch}
NUTS sampler (MC-NUTS).

From a computational standpoint, SAASGP is more resource-intensive than both BT-MLE and
SK-MLE. However, SAASGP tends to yield a more accurate model, particularly in scenarios
with limited data. For active learning workflow, it is crucial to have a reliable
surrogate model construction algorithm to faithfully model the underlying solution
surface, at least qualitatively, for optimization purposes. In order to quantify the
effectiveness of these aforementioned three model implementations to favor the best model
(i.e., BT-MLE, SK-MLE, and SAASGP), we focus on the following
metrics:

\begin{enumerate}
    \item[(a)] Correlation coefficient ($R^2$) on the holdout test dataset.
    \item[(b)] Coverage of the model, meaning how many actual values lie within
          the two standard deviations of the predicted values.
    \item[(c)] Relative computational cost to train the model.
\end{enumerate}

\begin{figure}
    \captionsetup{justification=centering}
    \includegraphics[clip, width=\columnwidth]{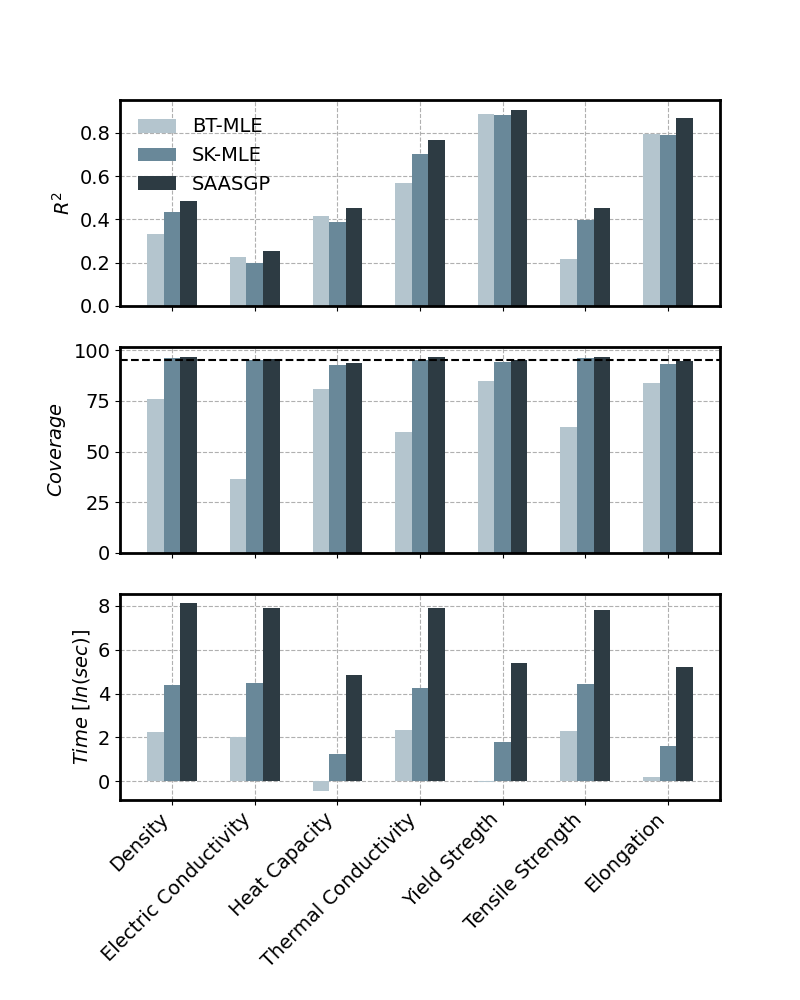}
    \caption{Comperative barplot of different GP model implementations. The top panel
        illustrates the $R^2$ value on the test set, the middle panel shows the coverage, and
        the third panel shows the time taken to train the model on a natural log scale (on a
        13th Gen Intel® Core™ i7-13700HX processor with 32 GB ram memory). We also
        show a black dotted line in the middle panel to indicate the 95\% coverage line to
        compare the quality of the uncertainty quantification capabilities of
        different GP model implementations.}
    \label{fig:barplot}
\end{figure}

The higher the $R^2$, the better the model predicts the underlying solution surface.
Similarly, from an uncertainty quantification perspective, a well-calibrated model
should have coverage that is close to 95\%. Despite significant improvements in
computer hardware and software, still the computational cost is a major hindrance to
apply Bayesian optimization to large-scale scientific problems. In figure \ref{fig:barplot},
we summarize the relative performance of these three implementations. The models are
trained five times with random 1500 data and tested on the rest of the data.

In our investigation, we found that the BT-MLE implementation performed consistently poorly
compared to the other two implementations. Interestingly, SK-MLE performed
as well as SAASGP, if not better, for our purpose.
Even though for some tasks, the $R^2$ is not as high as we would expect, we
still found them to be excellent for Bayesian optimization as the model can predict
the relative values quite faithfully, as evidenced by the coverage. Coverage for
both SK-MLE and SAASGP ranges around 90\% - 95\%, which
qualifies both to be very well-calibrated uncertainty quantification techniques.
However, if we observe the relative time to train the models, SK-MLE becomes
an automatic choice over the SAASGP. Considering the $R^2$ values,
coverage, and time, we used SK-MLE in our subsequent analysis. However,
SAASGP can be a powerful GP model for specific material science problems,
where the MAP-based model performs unsatisfactorily.

\subsection*{1-Dimensional Properties Optimization Results}

\begin{figure}
    \captionsetup{justification=centering}
    \centering
    \begin{NiceTabular}{cc}
        \Block[borders={top, tikz={solid, thick}}]{1-2}{}
        \Block[borders={bottom, tikz={densely dashed, thick}}]{1-2}{}
        \Block[borders={right,tikz={densely dashed, thick}}]{1-1}{}
        \subcaptionbox{Density Minimization \label{fig:density}} {\includegraphics[width=0.48\linewidth]{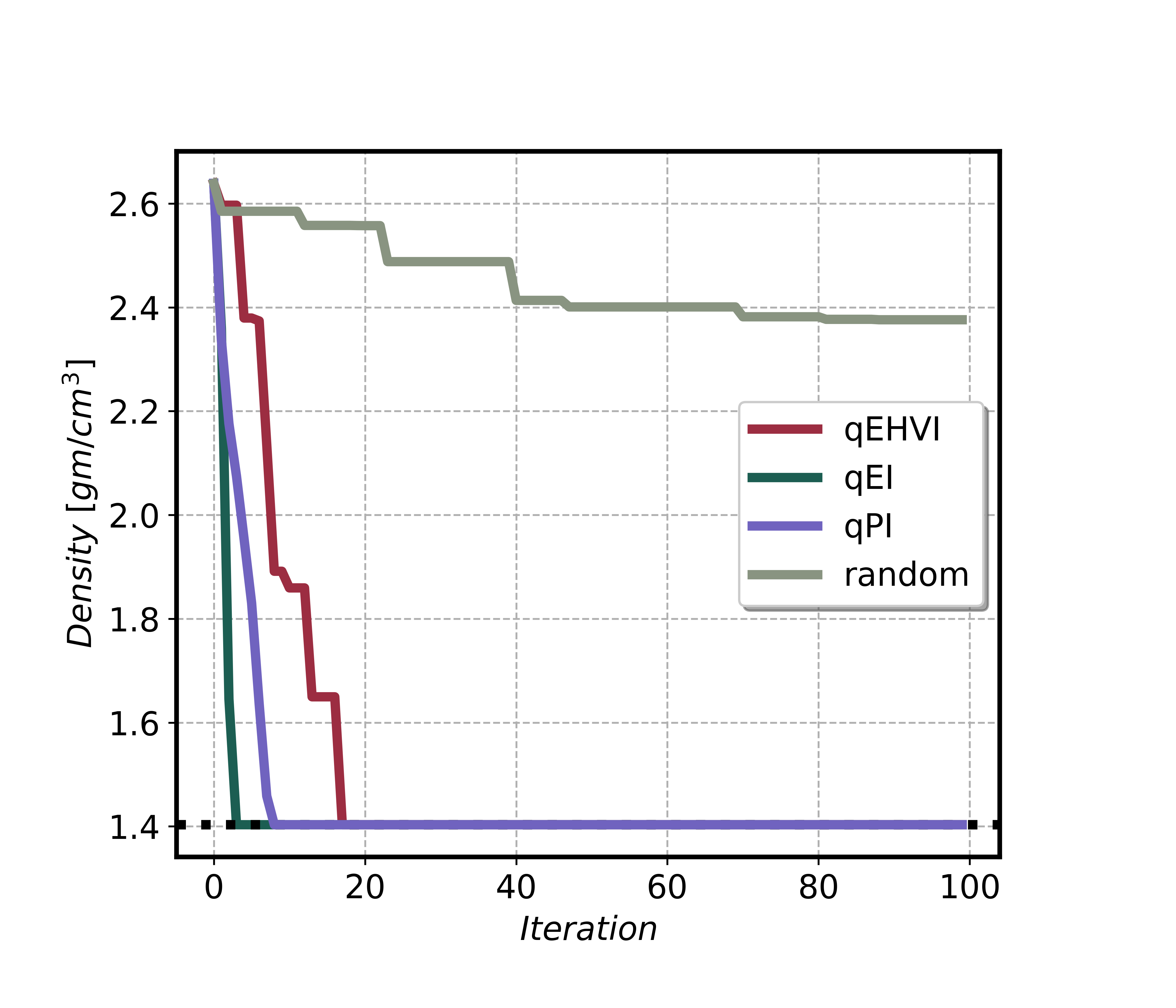}} &
        \subcaptionbox{Electric Conductivity Maximization \label{fig:electric}}{\includegraphics[width=0.48\linewidth]{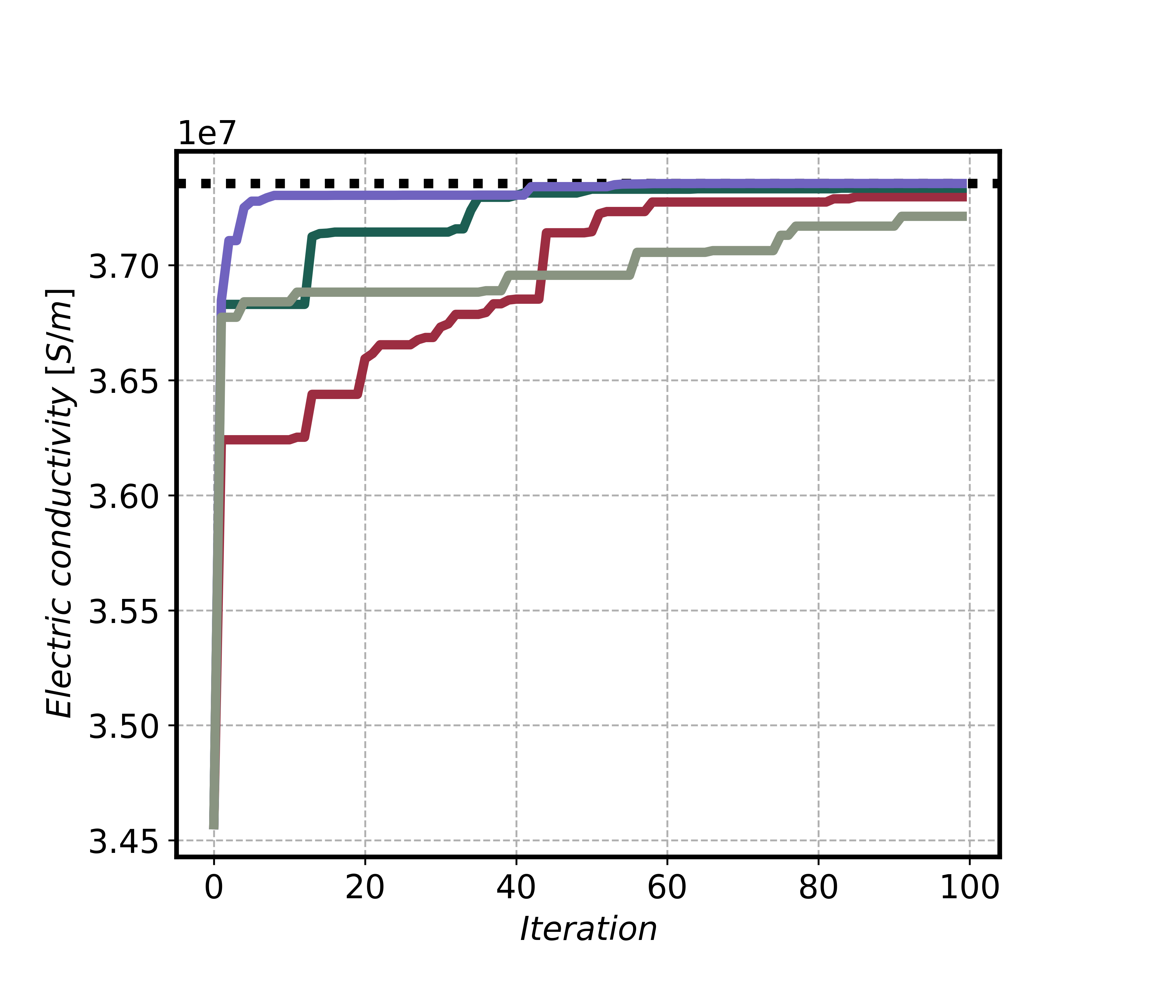}} \\
        \Block[borders={bottom, tikz={solid, thick}}]{1-2}{}
        \Block[borders={right,tikz={densely dashed, thick}}]{1-1}{}
        \subcaptionbox{Heat Capacity Maximization \label{fig:heat}} {\includegraphics[width=0.48\linewidth]{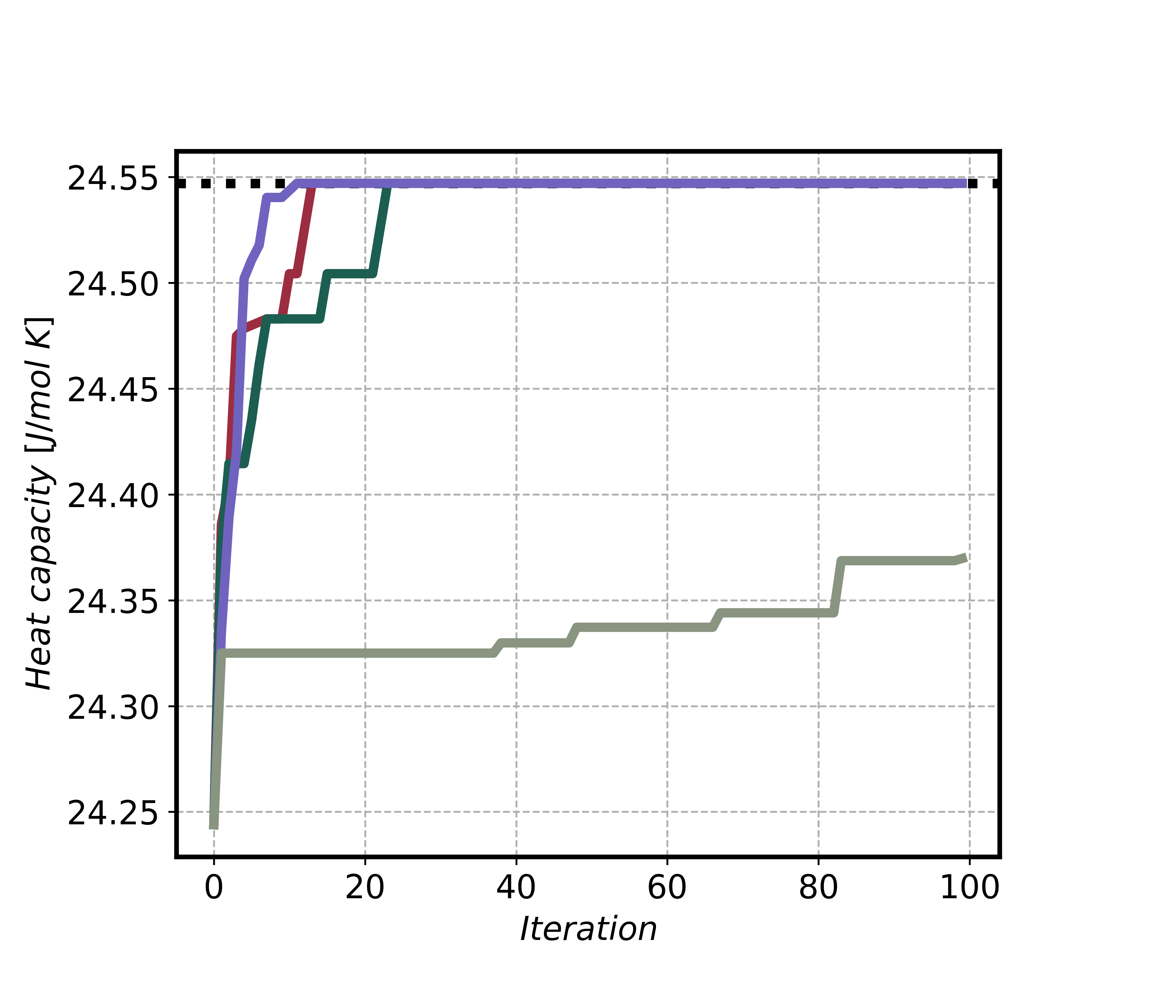}} &
        \subcaptionbox{Thermal Conductivity Maximization \label{fig:thermal}}{\includegraphics[width=0.48\linewidth]{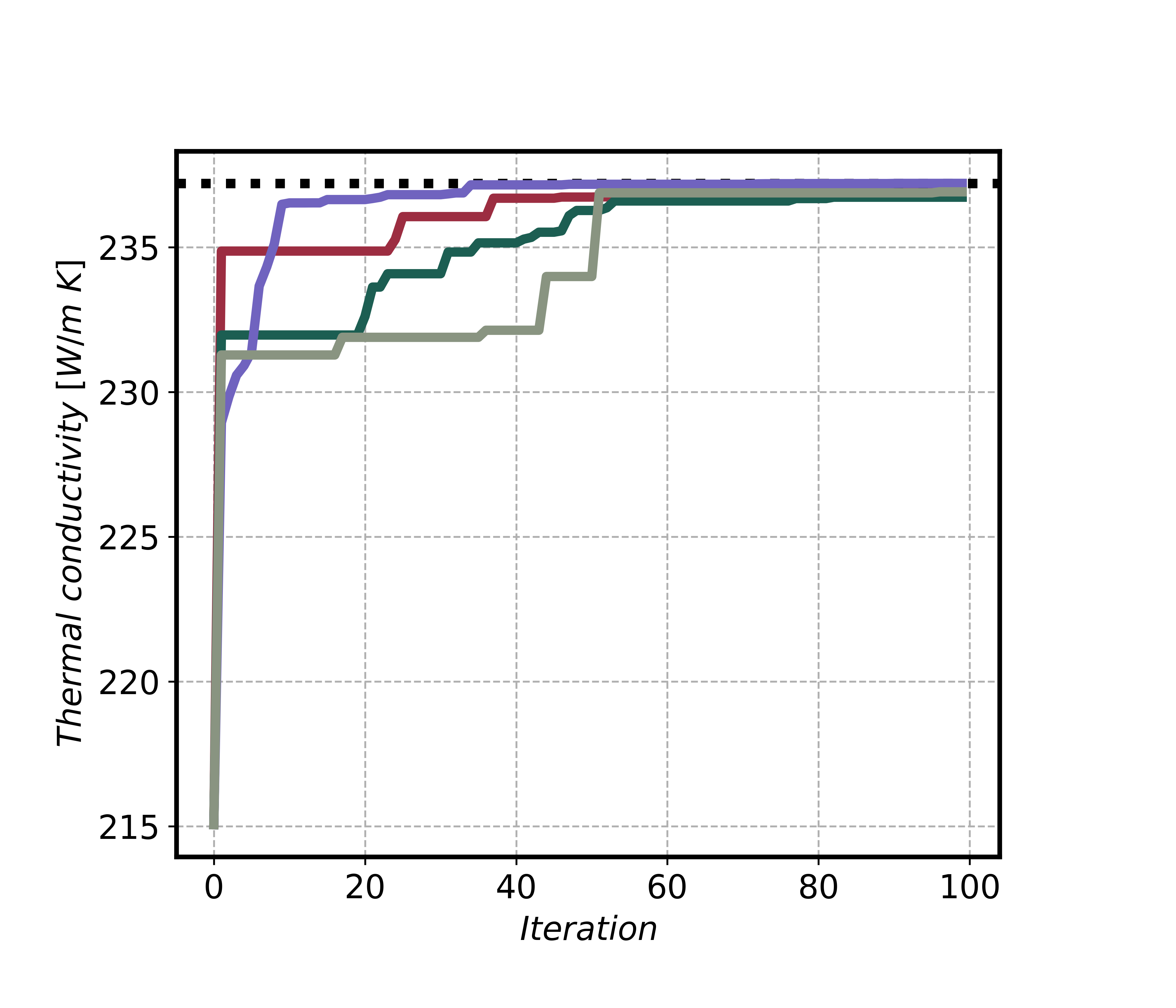}}
    \end{NiceTabular}
    \caption{Optimization profiles for 1-dimensional optimization of the computational
        properties with four different acquisition algorithms, i.e., qEHVI, qEI, qPI,
        and random. These profiles are generated using the average of five different
        runs with 25 initial random seed points.}
    \label{fig:one-d-comp-data}
\end{figure}

In Figure \ref{fig:one-d-comp-data}, we present the optimization profile for
various targets in the computational dataset for 1-dimensional optimization
problems. Bayesian optimization showed impressive performance
for density optimization in finding the optimum point (in this case, the minimum) by all three acquisition
functions, i.e., qEI, qPI, and qEHVI. Across all five runs, with
twenty-five seed points, the algorithm consistently reached the global minimum
(i.e., \(1.4\) \(\text{g.cm}^{-3}\)) in less than 20 iterations. This indicates
that to find the minimum point in a dataset of more than $13,000$ data points,
Bayesian optimization requires fewer than $50$ computations. In comparison, random
data acquisition could not reach the global minimum even once in the five trial
simulated runs.

For heat capacity optimization, the performance is similarly impressive in finding
the maximum, reaching it within $25$ iterations for all five runs, in stark
contrast to the poor performance of random data acquisition. In the case of
thermal conductivity, the average number of iterations to reach the global maximum
is about 40-50 for all three Bayesian optimization algorithms; however, random
data acquisition also performed similarly well for this problem, owing to the fact
that the data distribution is narrower and right-skewed (see supporting information)
it is easier to find materials that are close to the global maximum. For electric
conductivity, both qEI and qPI performed slightly better than the qEHVI. In this
instance, the performance of random acquisition was also similar to that of
Bayesian optimization.

These results highlight that the success of Bayesian optimization also depends
on the underlying data behavior. In scenarios with noise, a broad distribution
of the target property, or non-representative features, the performance of
Bayesian optimization may be less satisfactory. Effective application of
Bayesian optimization in realistic situations still greatly benefits from
domain knowledge, especially in designing the features or the surrogate model
that can make reliable predictions on unknown observations. From the one-dimensional
optimization of the computational data, we can conclude that the traditional
single-objective acquisition function is better suited for one-dimensional optimization
than the multiobjective acquisition function, such as qEHVI.

\begin{figure}
    \captionsetup{justification=centering}
    \centering
    \begin{NiceTabular}{cc}
        \Block[borders={top, tikz={solid, thick}}]{1-2}{}
        \Block[borders={bottom, tikz={densely dashed, thick}}]{1-2}{}
        \Block[borders={right,tikz={densely dashed, thick}}]{1-1}{}
        \subcaptionbox{Yield Strength Maximization \label{fig:yield}} {\includegraphics[width=0.48\linewidth]{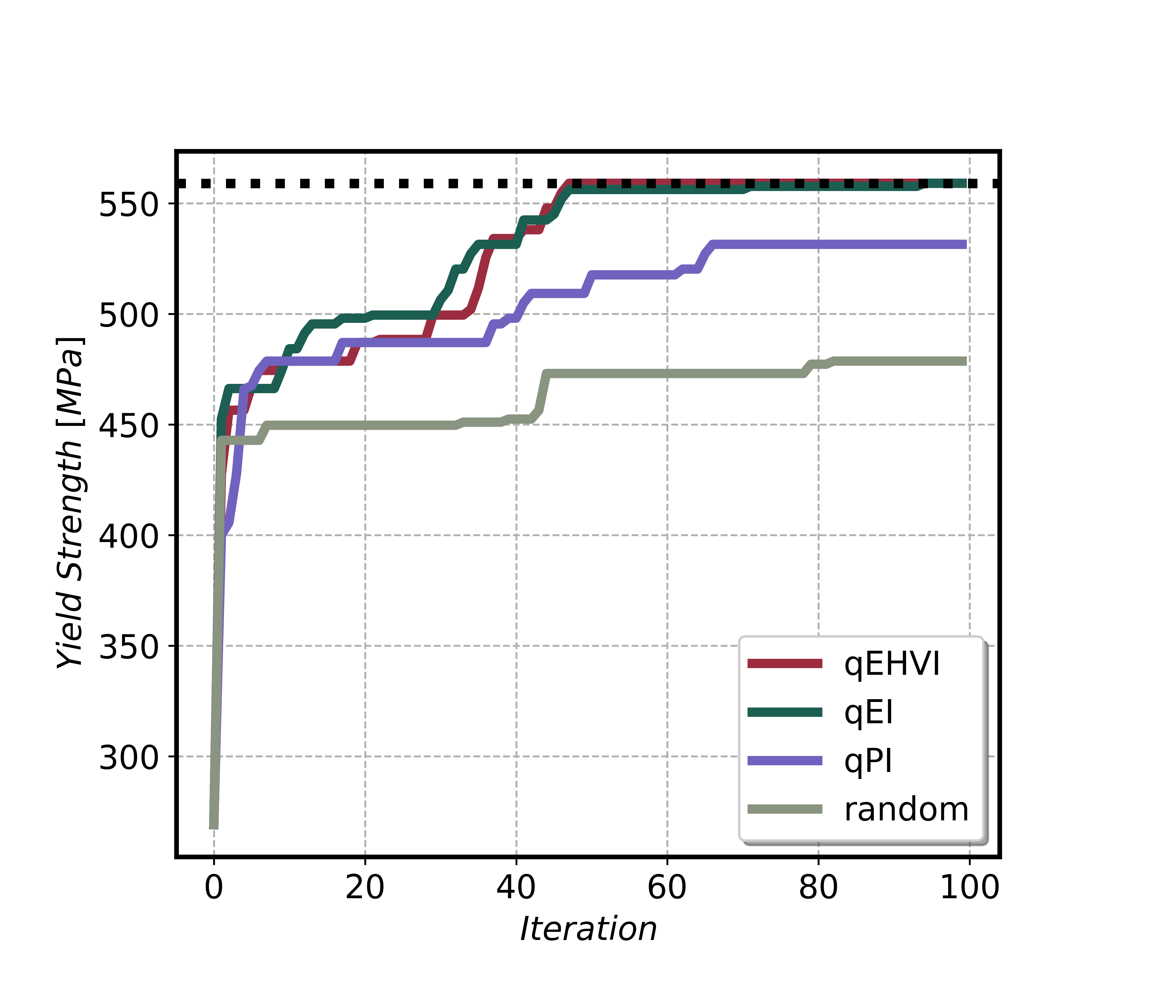}} &
        \subcaptionbox{Ultimate Tensile Strength Maximization \label{fig:tensile}}{\includegraphics[width=0.48\linewidth]{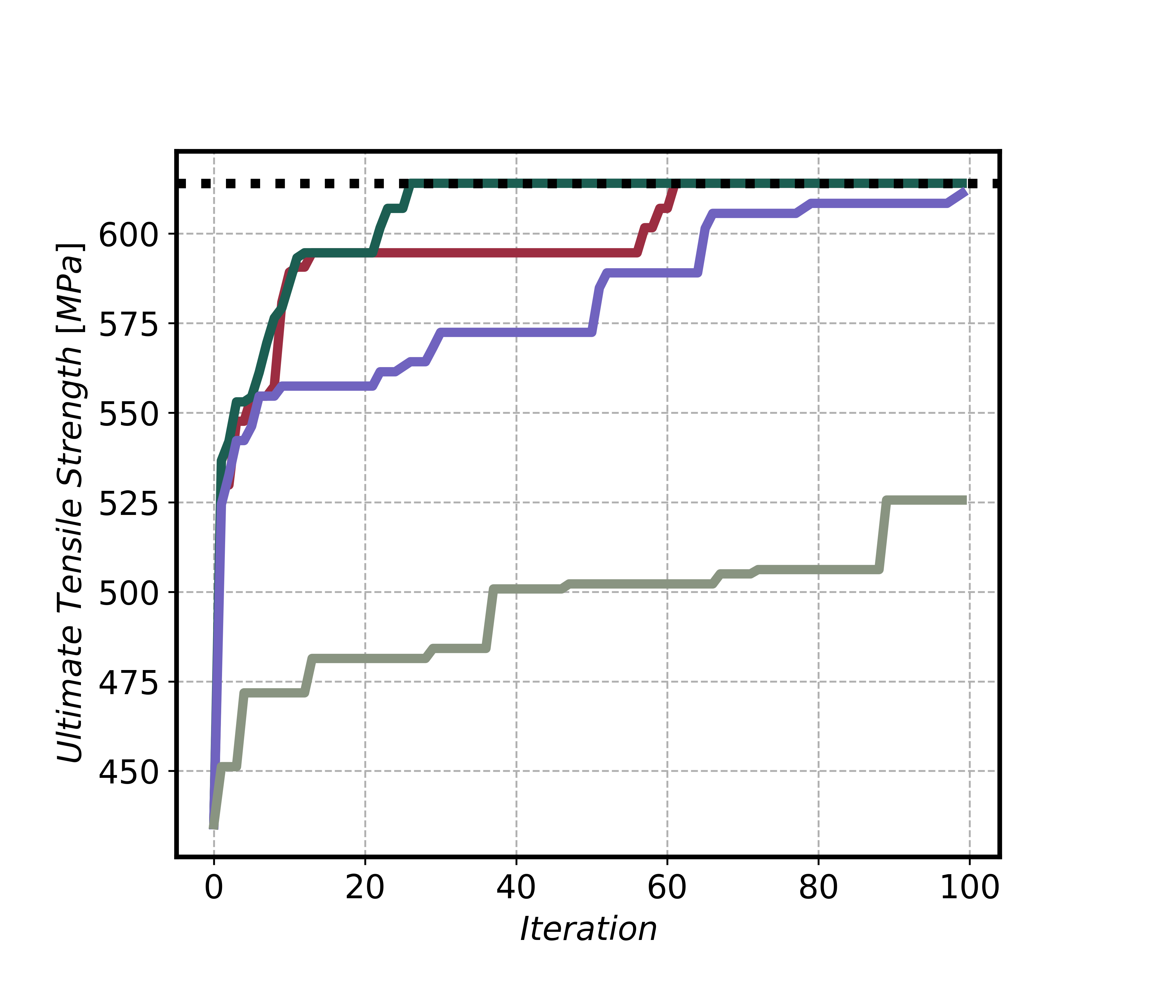}} \\
        \Block[borders={bottom, tikz={solid, thick}}]{1-2}{\subcaptionbox{Elongation Minimization \label{fig:elong}} {\includegraphics[width=0.48\linewidth]{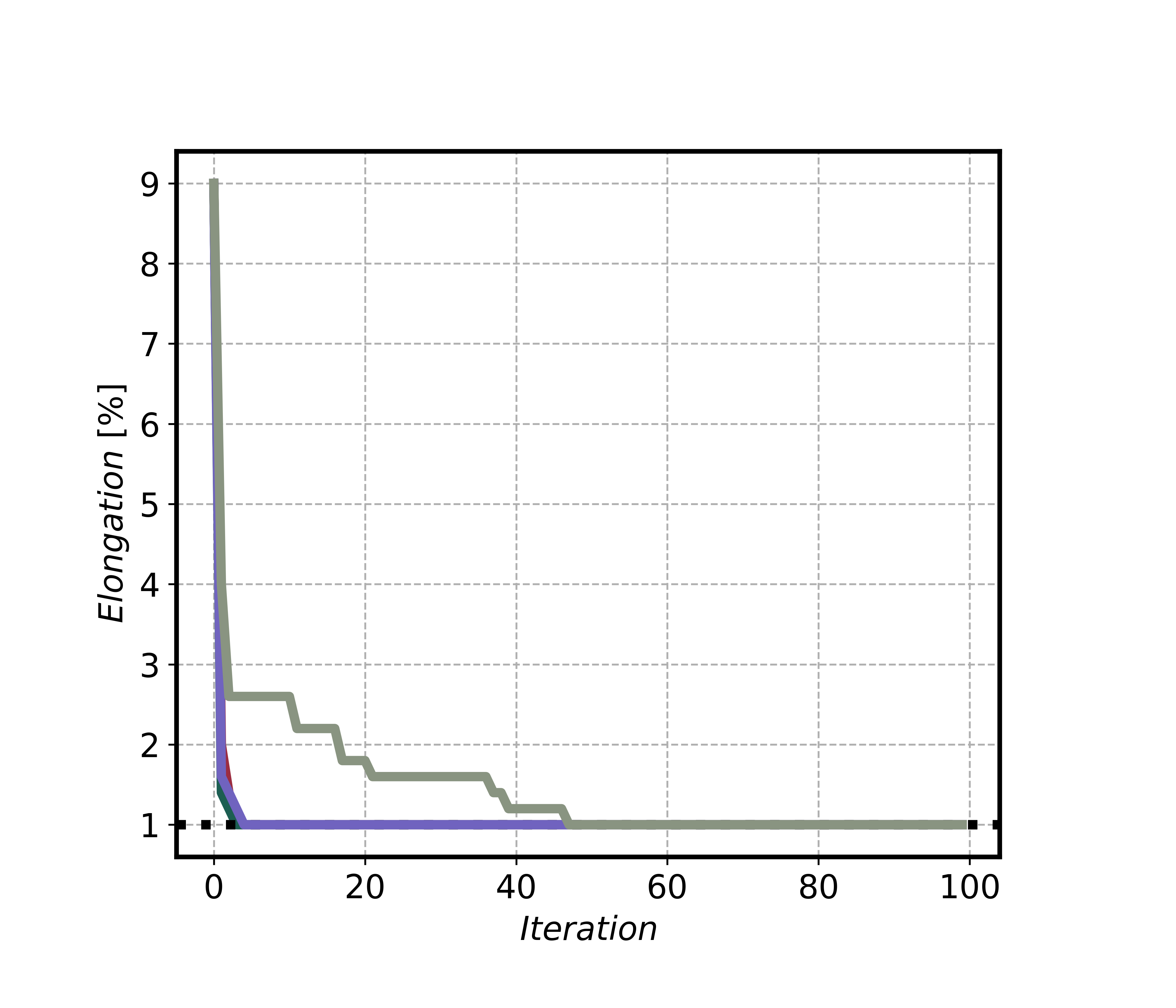}}}
    \end{NiceTabular}
    \caption{Optimization profiles for 1-dimensional optimization of the experimental
        properties with four different acquisition algorithms, i.e., qEHVI, qEI, qPI, and
        random. These profiles are generated using the average of five different runs
        with 25 initial random seed points.}
    \label{fig:one-d-exp-data}
\end{figure}

In Figure \ref{fig:one-d-exp-data}, we present the optimization profile for the
experimental dataset across various target properties, i.e., yield strength,
ultimate tensile strength, and elongation. The yield strength reaches its global
maximum (559 MPa) within 40 to 50 iterations for qEI and qEHVI but settles
at a local maximum (552 MPa) for qPI. Regarding ultimate tensile strength,
most trials achieve the global maximum within 20 to 60 iterations for qEI and qEHVI,
except for qPI, which reaches the global optimum on average at the 100th iteration.
All trials reach the global minimum for elongation in fewer than 15 iterations
for all three acquisition functions. In contrast, the random data acquisition failed
to reach the optimum for yield and ultimate tensile strength but managed
to reach the global minimum for elongation optimization at around 50 iterations.

Overall, all three acquisition functions performed better than the random acquisition
function across all tasks for both computational and experimental datasets, with qPI
being slightly worse than the other two. The reason for such underperformance is already
discussed above, i.e., qPI only considers the improvement over the best-known materials and
does not consider the relative improvement. Owing to this fact, PI often gets stuck
in a local optima. In contrast, the qEI and qEHVI are very comparable,
with qEI always performing slightly better at finding the global optimum within a minimal
number of evaluations. Regarding computational cost, none of the three acquisition functions
posed any significant bottleneck.

\subsection*{2-Dimensional Properties Optimization Results}

\begin{figure}
    \captionsetup{justification=centering}
    \centering
    \begin{NiceTabular}{cc}
        \Block[borders={top, tikz={solid, thick}}]{1-2}{}
        \Block[borders={bottom, tikz={densely dashed, thick}}]{1-2}{}
        \Block[borders={right,tikz={densely dashed, thick}}]{1-1}{}
        \subcaptionbox{Density Minimization and Heat Capacity Maximization \label{fig:density_heat}} {\includegraphics[width=0.48\linewidth]{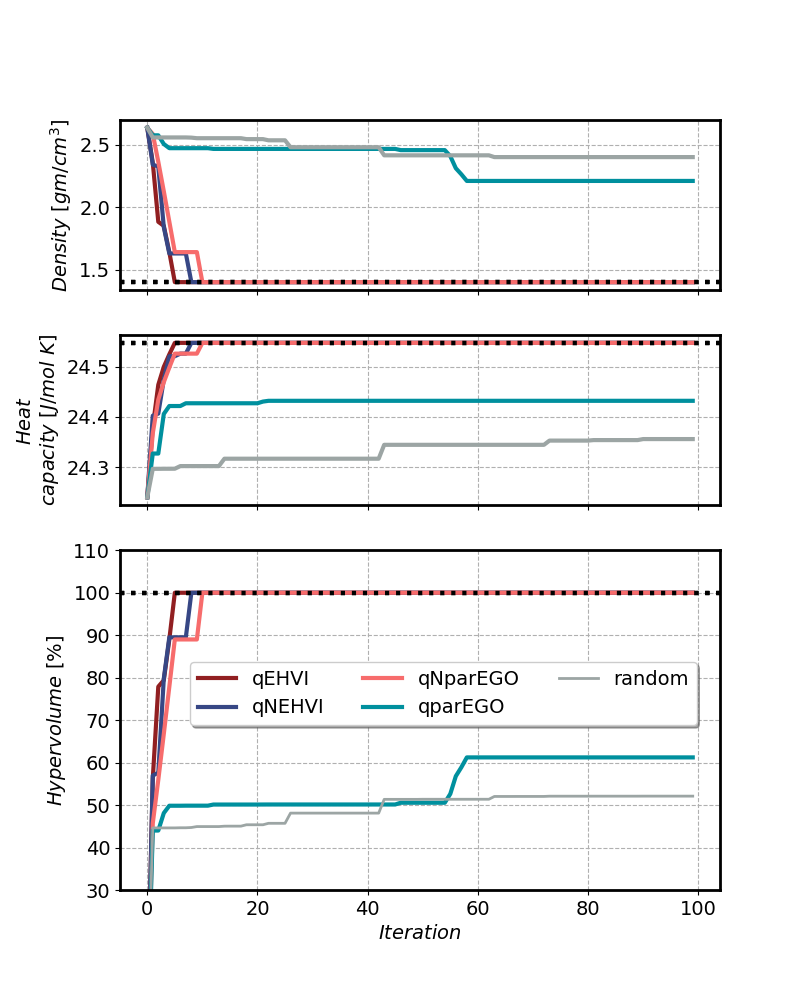}}              &
        \subcaptionbox{Density Minimization and Electric Conductivity Maximization \label{fig:density_electric}}{\includegraphics[width=0.48\linewidth]{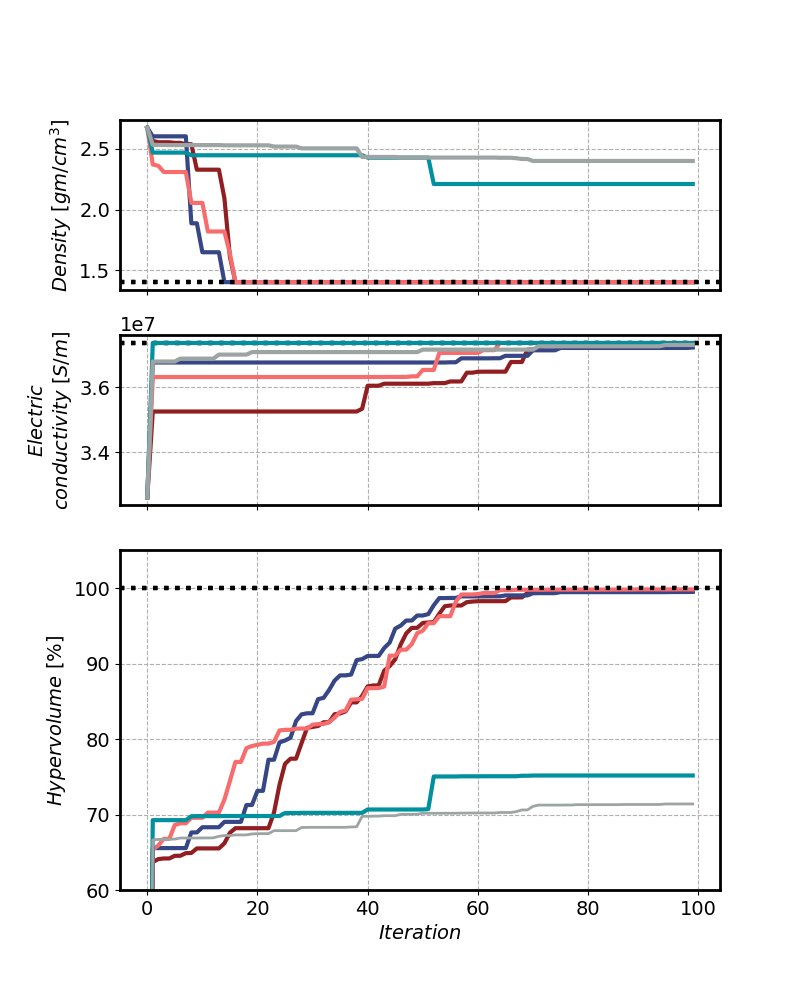}} \\
        \Block[borders={bottom, tikz={solid, thick}}]{1-2}{}
        \Block[borders={right,tikz={densely dashed, thick}}]{1-1}{}
        \subcaptionbox{Density Minimization and Thermal Conductivity Maximization \label{fig:density_thermal}} {\includegraphics[width=0.48\linewidth]{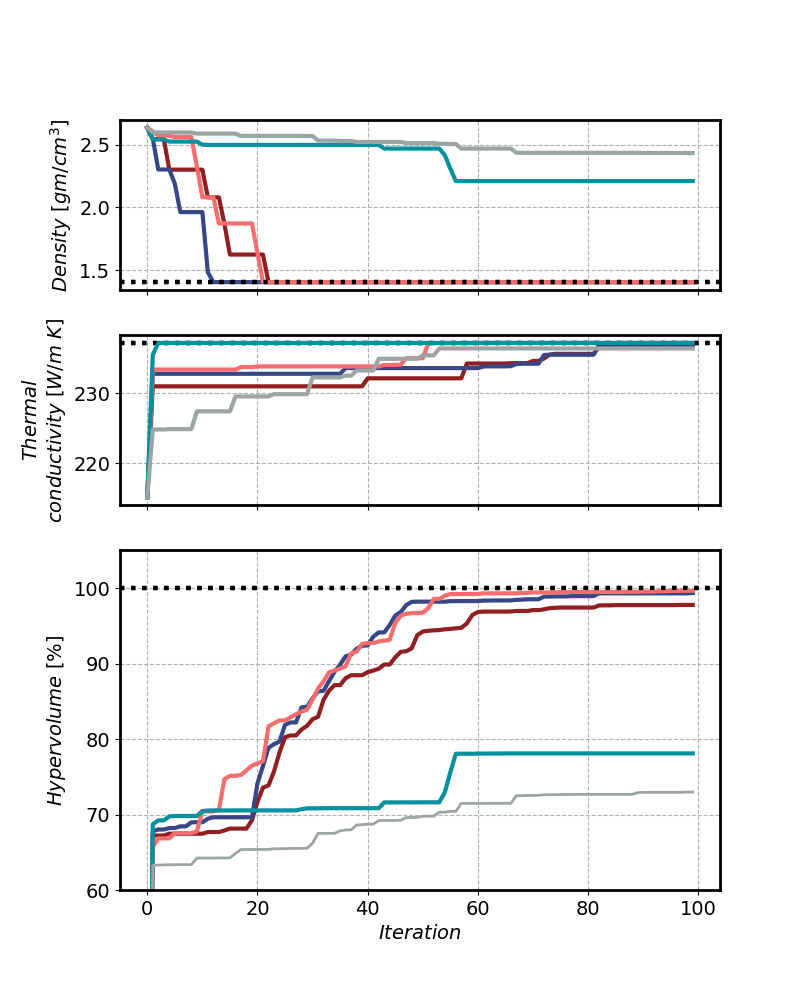}} &
        \subcaptionbox{Heat Capacity and Thermal Conductivity Maximization \label{fig:heat_thermal}}{\includegraphics[width=0.48\linewidth]{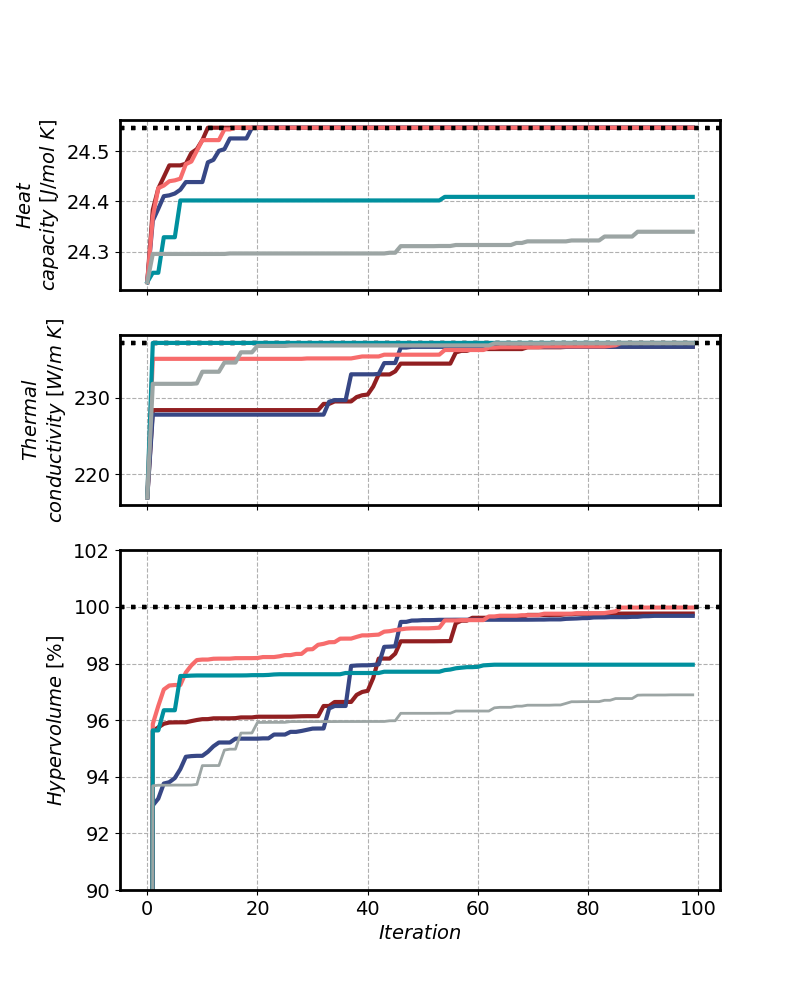}}
    \end{NiceTabular}
    \caption{Optimization profiles for 2-dimensional optimization of the computational
        properties with five different acquisition algorithms, i.e., qEHVI, qNEHVI, qparEGO,
        qNparEGO, and random. In each subfigures, the top two panel shows the optimization
        profile of the two properties during the multiobjective optimization run, while
        the third panel shows the hypervolume optimization profile. These profiles
        are generated using the average of five different runs with 25 initial random
        seed points.}
    \label{fig:two-d-comp-data}
\end{figure}

\begin{figure}
    \captionsetup{justification=centering}
    \centering
    \begin{NiceTabular}{cc}
        \Block[borders={top, tikz={solid, thick}}]{1-2}{}
        \Block[borders={bottom, tikz={densely dashed, thick}}]{1-2}{}
        \Block[borders={right,tikz={densely dashed, thick}}]{1-1}{}
        \subcaptionbox{Density Minimization and Electrical Conductivity Maximisation \label{fig:density_electric_pareto}} {\includegraphics[width=0.48\linewidth]{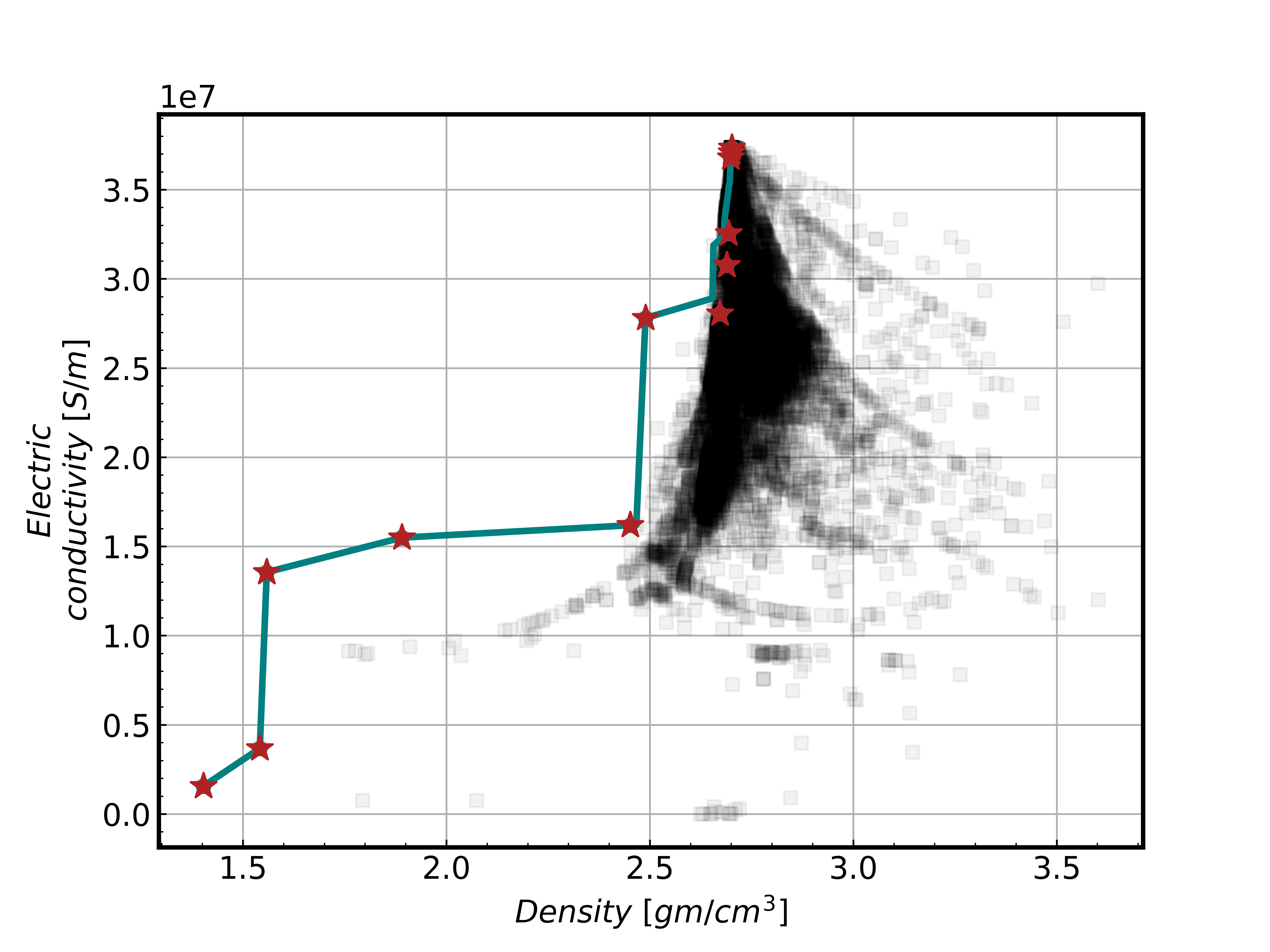}} &
        \subcaptionbox{Density Minimization and Thermal Conductivity Maximisation \label{fig:density_thermal_pareto}}{\includegraphics[width=0.48\linewidth]{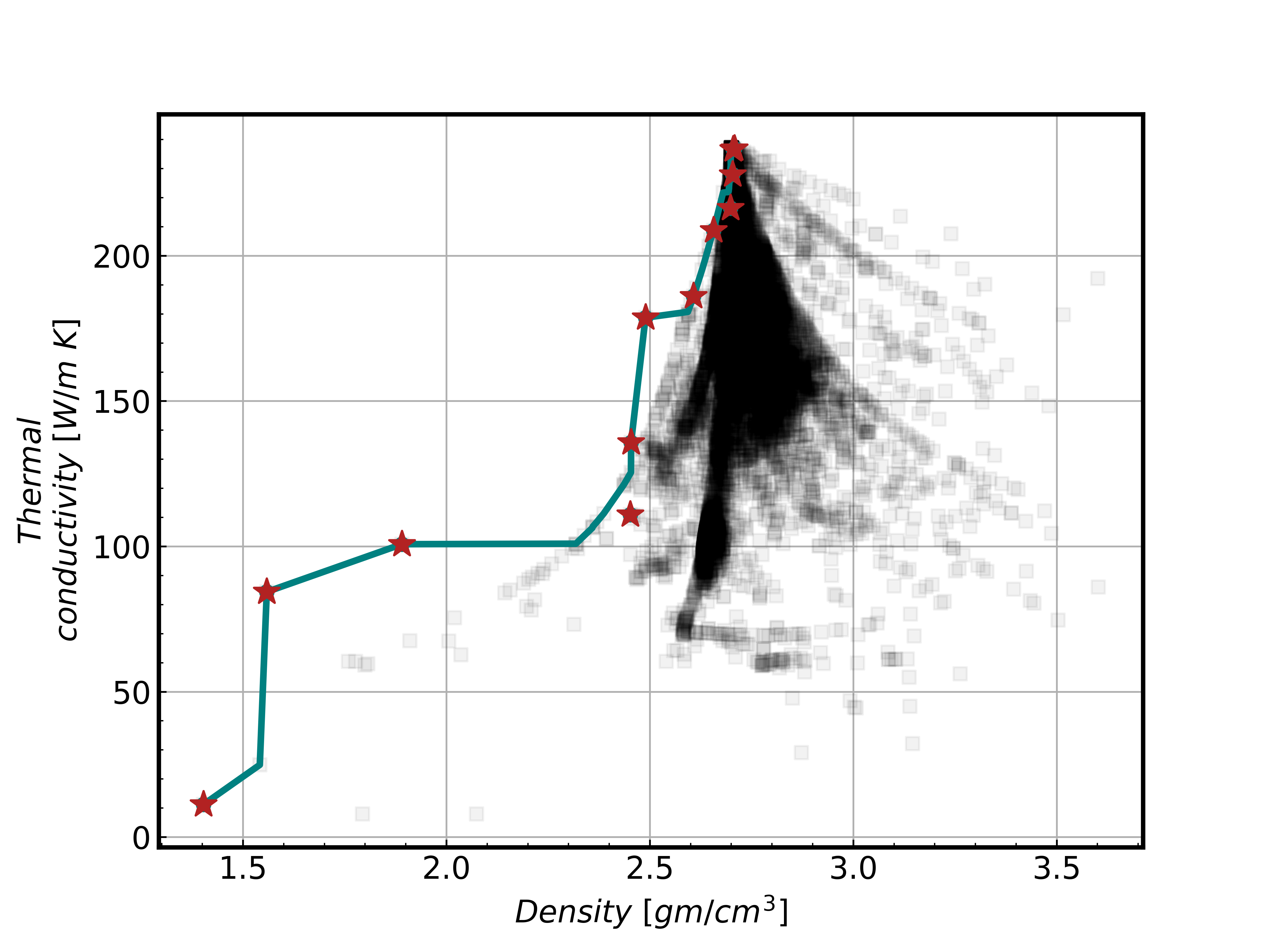}}         \\
        \Block[borders={bottom, tikz={solid, thick}}]{1-2}{\subcaptionbox{Ultimate Tensile Strength Maximization and Elongation Minimization \label{fig:tensile_elong_pareto}} {\includegraphics[width=0.48\linewidth]{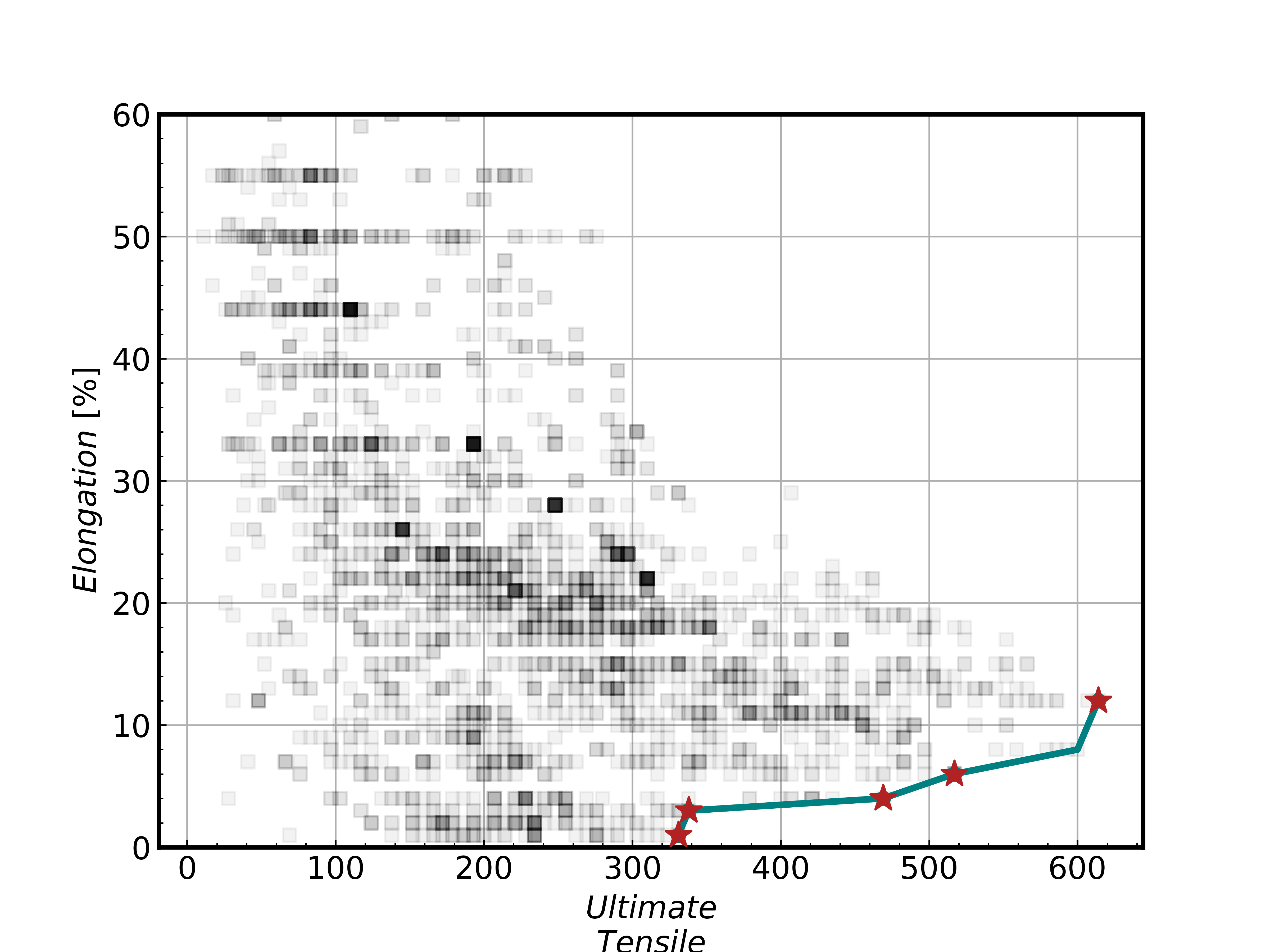}}}
    \end{NiceTabular}
    \caption{Pareto front of the selected runs for some selected 2-dimensional
        optimization. The global Pareto front is shown as a teal line while the Pareto
        front obtained at the end of the simulations is shown as a red marker ($\star$).
        The black squares represent all the data in the database (a) Density Minimization
        and Electrical Conductivity Maximization, (b) Density Minimization and Thermal
        Conductivity Maximization, and (c) Elongation Minimization and Ultimate Tensile
        Strength Maximization. For computational data ((a) \& (b)), we use qEHVI, and
        for experimental data ((c)), we use qNEHVI.}
    \label{fig:two-d-pareto}
\end{figure}

Next, we explore the two-dimensional optimization profile for computational
problems (refer to Figure \ref{fig:two-d-comp-data}). Focusing on novel materials
discovery, which typically involves finding low-density materials with high heat
capacity, thermal conductivity, and electrical conductivity, we aimed to minimize
density while maximizing heat capacity, electrical conductivity, or thermal
conductivity in two-dimensional optimizations. Additionally, we designed a
problem to maximize heat capacity and thermal conductivity. In all four
simulations, Bayesian Optimization (BO) outperformed random data acquisition,
sometimes reaching the global maximum hypervolume within 50 iterations (starting
with 25 seed points). To illustrate the quality of the acquired data points, we
also present Pareto fronts for three runs with the best acquisition function
identified in this work across three different optimization problems in Figure
\ref{fig:two-d-pareto}. As the objective was to maximize the hypervolume, the
algorithm effectively identified points forming the basis of the Pareto front.
However, extending the run duration or modifying the objective function could
allow further exploration of specific regions in space, showcasing constrained
BO as a potent tool.

In all three density minimization problems (with maximization
of heat capacity, thermal conductivity, or electric conductivity), only qparEGO
acquisition function performed poorly (only slightly better than random data
acquisition); however, the other three, i.e., qEHVI, qNEHVI, and qNparEGO, showed
impressive results, e.g., reaching the global maximum hypervolume within 20 for
density minimization and heat capacity maximization problem. For the fourth problem,
i.e., heat capacity and thermal conductivity maximization, all four acquisition
functions, and random data acquisition performed exceptionally well, reaching
$>~95\%$ hypervolume within 40 iterations. If we observe the one-dimensional
optimization profile on the top panels of the associated subplots, it is quite
evident that qparEGO, being a scalarization-based acquisition function, can
comfortably optimize one of the objective functions, but falls short on the
other objective function, which could be attributed to the poor choice of the
scalarization function that most likely favored one objective over the other.
Selection of suitable scalarization function is an active area of research.
By taking the uncertainty into account, the parEGO
formulation (qNparEGO) can perform as well as EHVI-based methods, if not better.
However, there is negligible performance difference between the qEHVI and its noisy
variant (qNEHVI). It is of note that both the noisy variants, i.e., qNEHVI and
qNparEGO required more time to compute the acquisition function values than
their non-noisy counterparts. On average, they took 5-6 times more compute across
all iterations (iteration time is smaller in the beginning of the simulations and
slowly rises as we include more training samples to the active learning cycle).

For the experimental dataset, we maximized yield strength
and ultimate tensile strength in one simulation, observing both the noisy variant
performing quite well, with the qEHVI being very close to the maximum hypervolume
and qparEGO being no better than the random data acquisition function.
We maximized yield strength and ultimate tensile strength in two separate
simulations while minimizing elongation. Although none reached the optimal point
within our set 50 iterations, they achieved more than 90\% of the maximum
hypervolume within a budget of 50 experiments (excluding 25 seed points).
Similar to the computational dataset, qparEGO performed very poorly, barely
surpassing the random data acquisition function. In contrast, qEHVI, qNEHVI, and
qNparEGO, all three, performed objectively better than the random data acquisition;
however, qNEHVI turned out to be superior, owing to the fact that experimental
data contains significant noise. The Pareto front for the simulation of
minimization of elongation and maximization of ultimate tensile strength demonstrates
its capability to identify almost all points on the global Pareto front.
From this comparative analysis of computational and experimental datasets, it becomes
evident that the qEHVI algorithm more readily identifies optima in the computational
dataset with minimal computational cost, likely due to its noise-free nature.
In contrast, qNEHVI performed better for the the experimental data optimization
problems, owing to the fact that the data collected from experiments are subjected to
conditions and measurement apparatus fluctuations, and likely contains more noise
than the computational data.

\subsection*{3-Dimensional Properties Optimization Results}

We also performed two 3-dimensional optimizations utilizing the computational dataset:

\begin{enumerate}
    \item Density minimization along with heat capacity and electric conductivity maximization.
    \item Density minimization along with heat capacity and thermal conductivity maximization.
\end{enumerate}

Due to limited working computational memory (32 GB RAM), we could only run 40
iterations before hitting the working memory bottleneck. Similar to
2-dimensional optimization problems, both problems are satisfactorily optimized
towards the global Pareto frontier by qEHVI, qNEHVI, and qNparEGO. Based on this
analysis, given better computational resources, we can potentially optimize
multi-objective problems involving objectives ($ m \ge 3 $) with this framework.

\subsection*{Multiple Data Acquisition Strategy}
Finally, we also ran simulations where, instead of choosing one candidate point
at each iteration, we chose multiple candidate points, e.g., two or four points,
at each iteration. This could be preferable when limited time is available for
exploration/exploitation, but facilities to run multiple
experiments/computations are also available. By acquiring multiple candidate
data points, the number of iterations is reduced significantly to reach a
similar performance as a single data acquisition cycle. However, it scales sparingly;
for instance, by acquiring two candidate points, one would not
necessarily expect the number of iterations to be halved. In addition, owing to
the intensive nature of Monte-Carlo simulations, we can only acquire 2 or 4 candidates
successfully at each iteration before computational time and memory becomes an issue.

\section{Conclusion}
Developing rapid and reliable methods for multi-objective property optimization
in materials science is essential in the era of the Materials Genome
Initiative (MGI). In this context, we have demonstrated that q-expected
hypervolume improvement (qEHVI), a hypervolume-based acquisition function, is
highly effective for optimizing multi-component alloy systems with one-, two-
and three-dimensional objectives. We also discovered that the Pareto frontier
identified by this method after 50 iterations closely aligns with the global
Pareto front. However, this model presupposes prior knowledge of the
observations for candidate selection (discrete design space). In realistic
scenarios, the necessary observations for candidate selection are often
need to be made available, or their acquisition requires simplifying assumptions about
the search space. To address this limitation for numerical features, a more suitable
approach would involve selecting candidates from a continuous search space.
Furthermore, to render this optimization framework practically feasible, at
least for simulations, computing the properties with on-the-fly computations
after each iteration would be advantageous. We aim to make this framework
user-friendly for discovering novel materials. To achieve this,
we persistently adapting this method to a continuous search space
with on-the-fly simulation capabilities. Additionally, we aim to encapsulate
it in a user-friendly graphical user interface (GUI) for the materials science
community in the near future. In conclusion, we can summarise the pivotal
characteristics of this study:
\begin{itemize}
    \item Singular efficiency in single-property optimization: Demonstrating
          remarkable efficiency, it identifies the optimal data within a dataset of
          Thirteen thousand instances in fewer than 75 evaluations.
    \item Multi-objective optimization prowess: For multi-objective optimization,
          our algorithm discerns a well-defined Pareto front within 70-75 iterations,
          closely aligned with the global Pareto front by achieving a hyper-volume of
          90\% to 100\% of the maximum attainable.
    \item Enhanced candidate recommendation: It excels in recommending multiple
          candidates at each iteration, substantially expediting the discovery of novel
          alloys where parallel experimental or computational facilities are available.
    \item Rigorous construction of Surrogate model: Depending on the data quality
          and quantity, this framework discusses the best practices for constructing surrogate
          models that maximize the uncertainty quantification capability alongside
          minimizing computational cost.
\end{itemize}

\begin{figure}
    \captionsetup{justification=centering}
    \centering
    \begin{NiceTabular}{cc}
        \Block[borders={top, tikz={solid, thick}}]{1-2}{}
        \Block[borders={bottom, tikz={densely dashed, thick}}]{1-2}{}
        \Block[borders={right,tikz={densely dashed, thick}}]{1-1}{}
        \subcaptionbox{Yield Strength and Ultimate Tensile Strength Maximization \label{fig:yield_tensile}} {\includegraphics[width=0.48\linewidth]{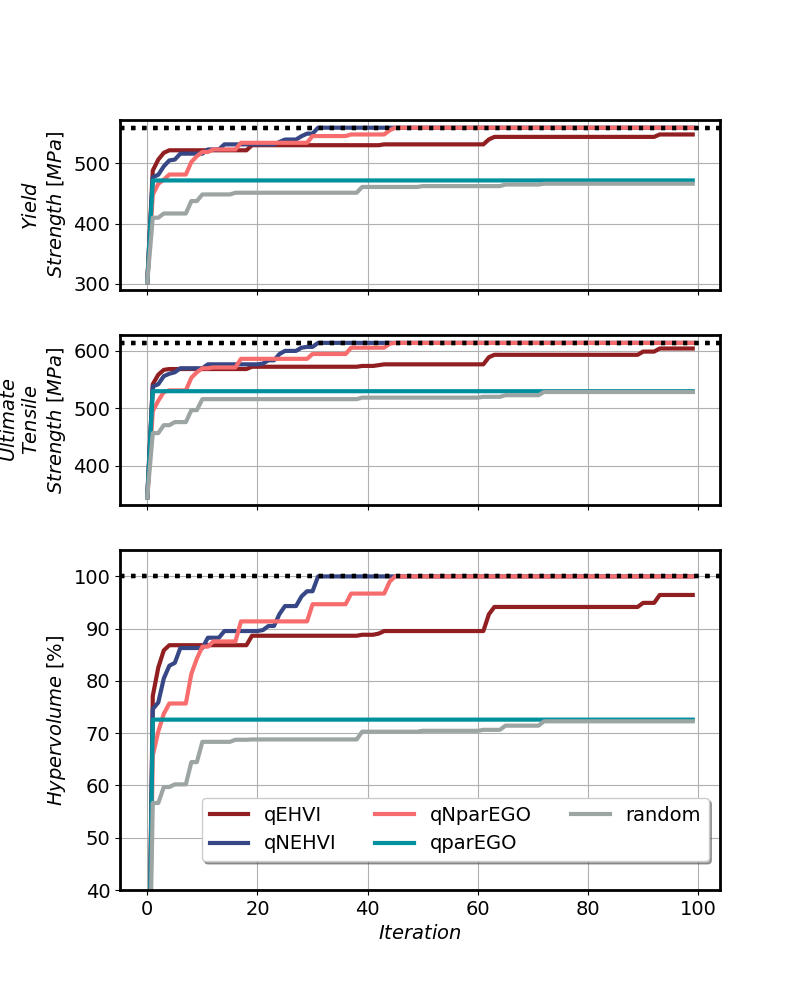}} &
        \subcaptionbox{Yield Strength Maximization and Elongation Minimization \label{fig:yield_elong}}{\includegraphics[width=0.48\linewidth]{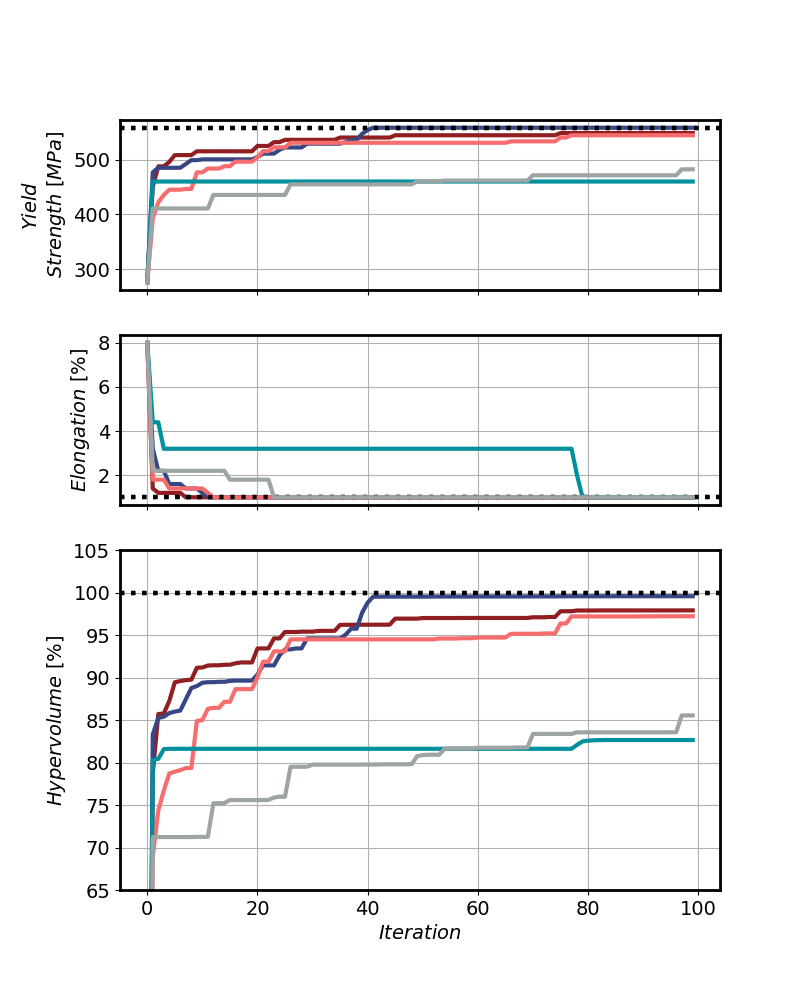}}          \\
        \Block[borders={bottom, tikz={solid, thick}}]{1-2}{\subcaptionbox{Ultimate Tensile Strength Maximization and Elongation Minimization \label{fig:tensile_elong}} {\includegraphics[width=0.48\linewidth]{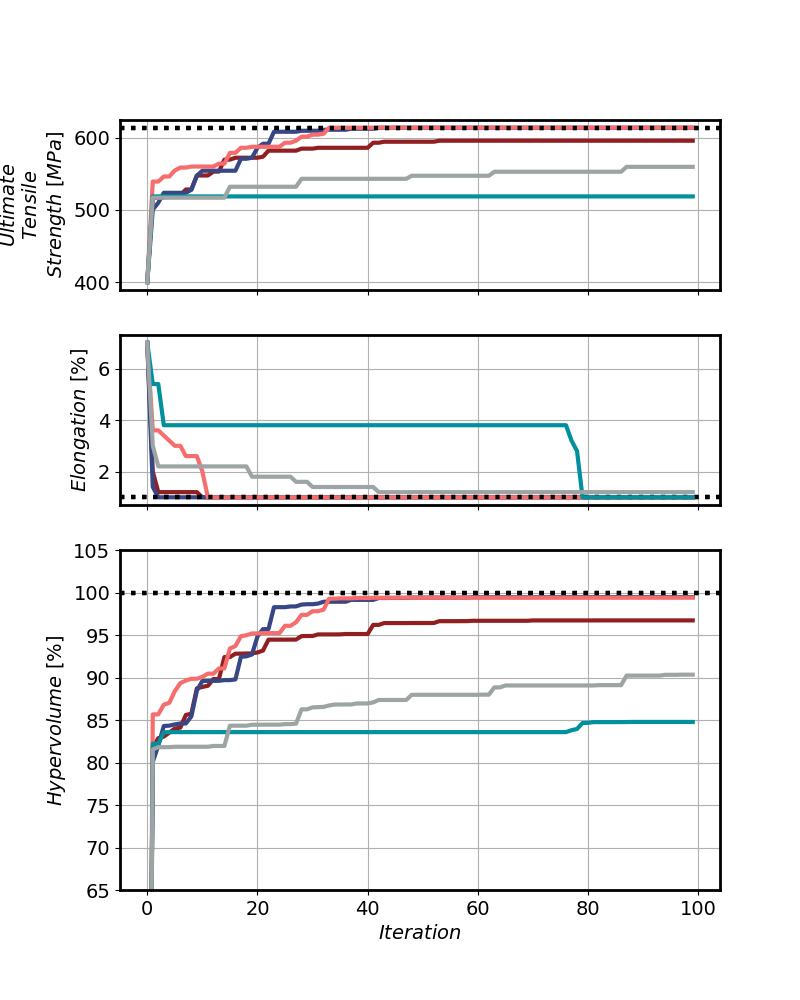}}}
    \end{NiceTabular}
    \caption{Optimization profiles for 2-dimensional optimization of the experimental
        properties with five different acquisition algorithms, i.e., qEHVI, qNEHVI, qparEGO,
        qNparEGO, and random. In each subfigures, the top two panel shows the optimization
        profile of the two properties during the multiobjective optimization run, while the
        third panel shows the hypervolume optimization profile. These profiles are
        generated using the average of five different runs with 25 initial random seed points.}
    \label{fig:two-d-exp-data}
\end{figure}

\begin{figure}
    \captionsetup{justification=centering}
    \centering
    \begin{NiceTabular}{cc}
        \Block[borders={top, tikz={solid, thick}}]{1-2}{}
        \Block[borders={bottom, tikz={solid, thick}}]{1-2}{}
        \Block[borders={right,tikz={densely dashed, thick}}]{1-1}{}
        \subcaptionbox{Density Minimization and Heat Capacity \& Electric Conductivity Maximization \label{fig:density_heat_electric}} {\includegraphics[width=0.48\linewidth]{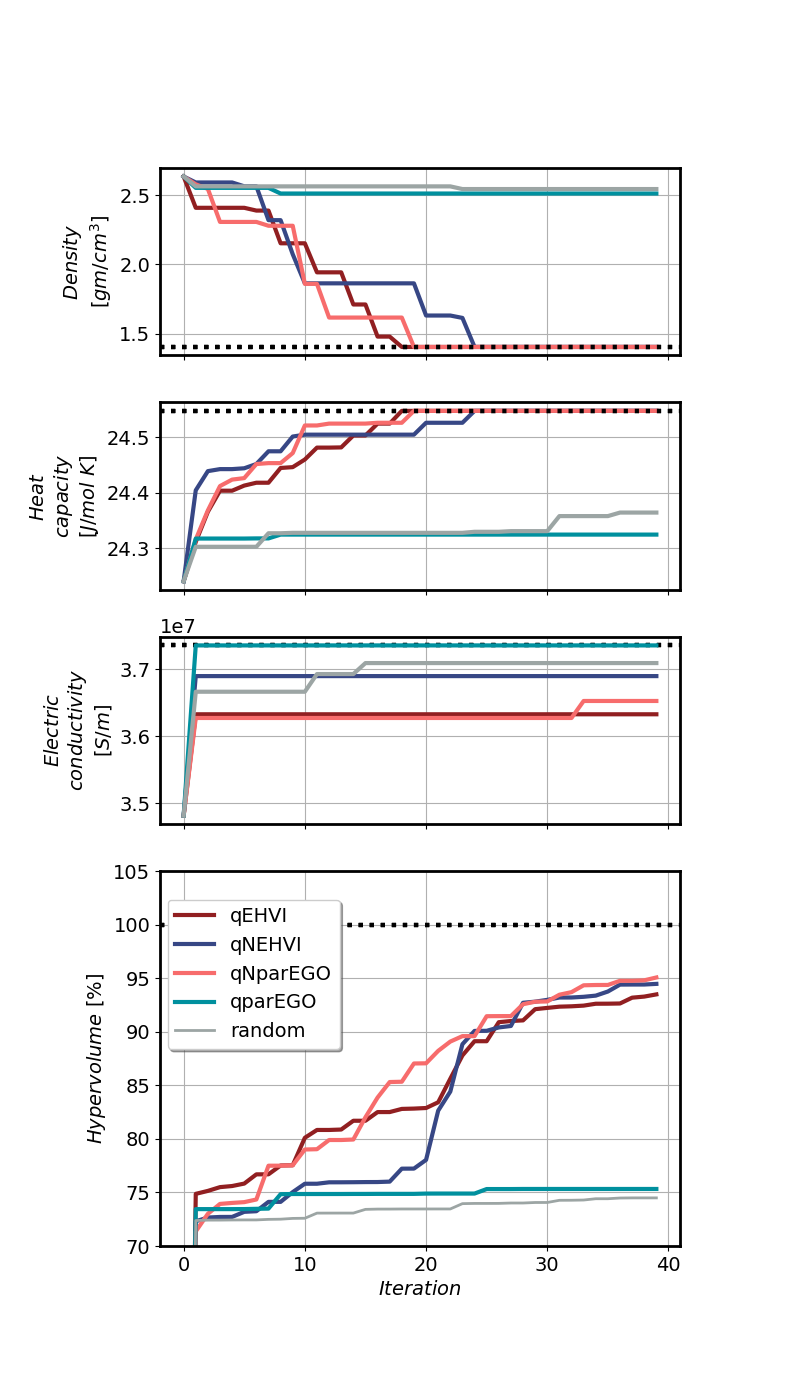}} &
        \subcaptionbox{Density Minimization and Heat Capacity \& Thermal Conductivity Maximization \label{fig:density_heat_thermal}}{\includegraphics[width=0.48\linewidth]{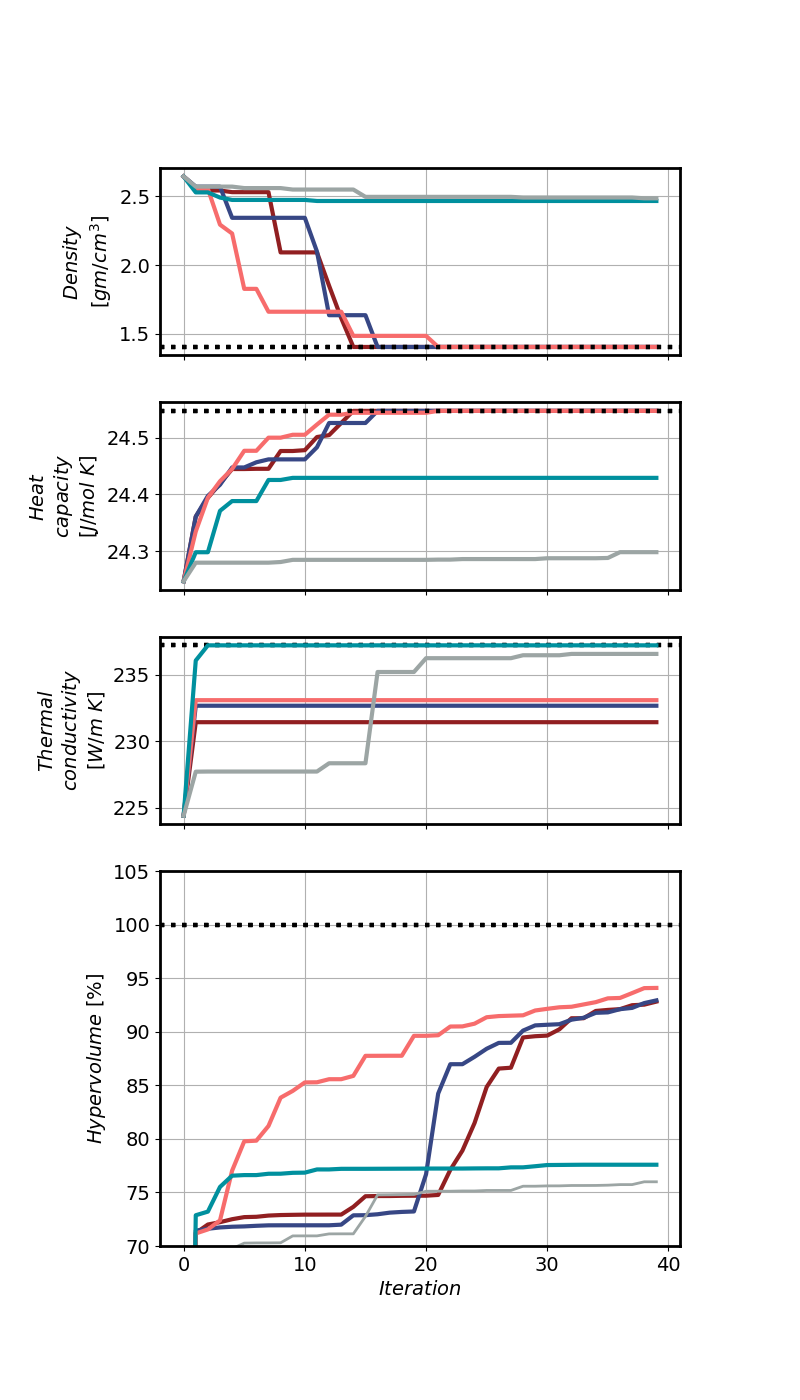}}
    \end{NiceTabular}
    \caption{Optimization profiles for 3-dimensional optimization of the computational
        properties with five different acquisition algorithms, i.e., qEHVI, qNEHVI, qparEGO,
        qNparEGO, and random. In each subfigures, the top three panel shows the optimization
        profile of the three properties during the multiobjective optimization run, while the
        fourth panel shows the hypervolume optimization profile. These profiles are generated
        using the average of five different runs with 25 initial random seed points.}
    \label{fig:three-d-comp-data}
\end{figure}

\section*{Data And Code Availability Statement}
The code is available at https://github.com/mamunm/AlloyOpt. Trained Bayesian optimisation (BO) processes
based on Active Learning (AL) principles, and the data underpinning the conclusions
of this study, are accessible from the corresponding author upon a reasonable request.

\bibliography{references}

\end{document}


\nolinenumbers

\title{Supporting Documents for Accelerated Development of Multi-component Alloys in Discrete Design Space Using Bayesian Multi-Objective Optimization}

\author[1]{Osman Mamun}
\author[2]{Markus Bause}
\author[1,2,*]{Bhuiyan Shameem Mahmood Ebna Hai}

\affil[1]{Fehrmann MaterialsX GmbH - Fehrmann Tech Group, Hamburg, Germany}
\affil[2]{Helmut Schmidt University - University of the Federal Armed Forces Hamburg, Germany}
\affil[*]{Corresponding author: shameem.ebna.hai@fehrmann-materialsx.com}


\maketitle

\renewcommand{\figurename}{Supporting Figure}

\section{Appendix A}

\begin{figure}
    \centering
    \captionsetup{justification=centering}
    \begin{subfigure}[b]{0.49\textwidth}
        \centering
        \includegraphics[width=\textwidth]{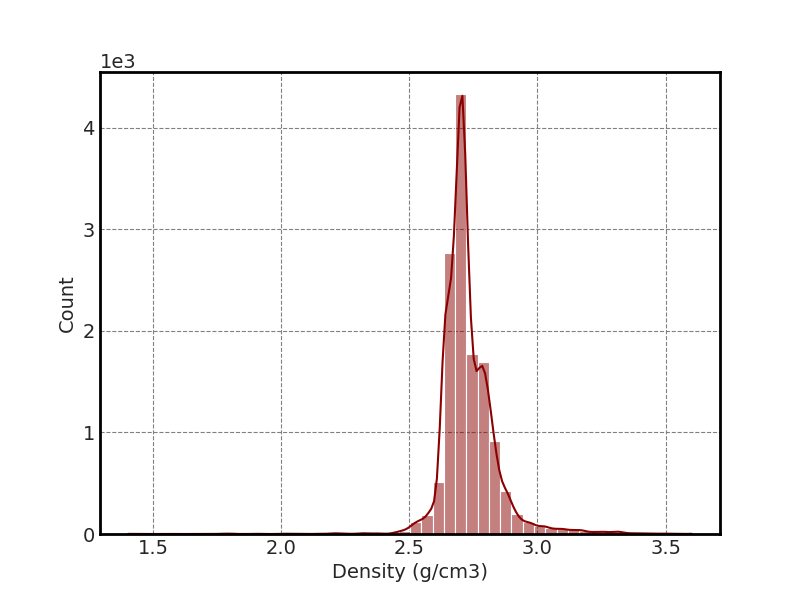}
        \caption{Density}
        \label{fig:density_dist}
    \end{subfigure}
    \hfill
    \begin{subfigure}[b]{0.49\textwidth}
        \centering
        \includegraphics[width=\textwidth]{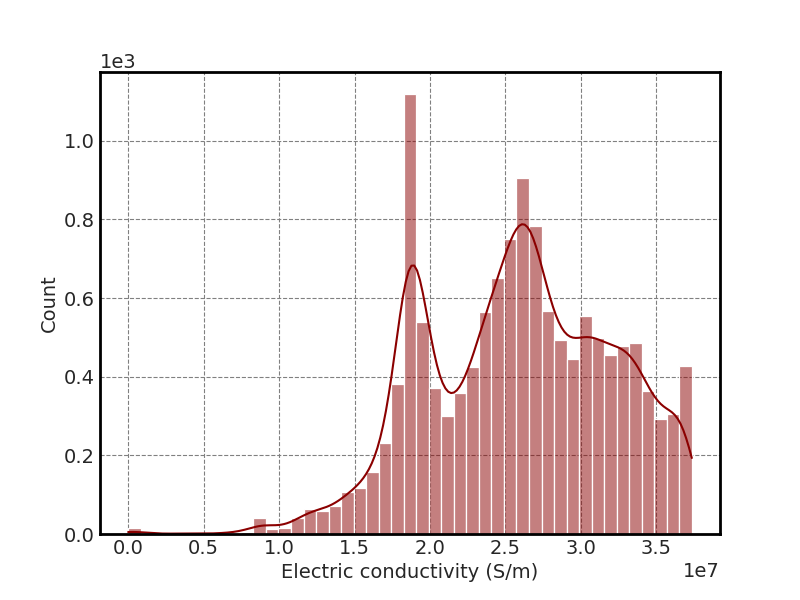}
        \caption{Electrical Conductivity}
        \label{fig:elec_cond_dist}
    \end{subfigure}
    \hfill
    \begin{subfigure}[b]{0.49\textwidth}
        \centering
        \includegraphics[width=\textwidth]{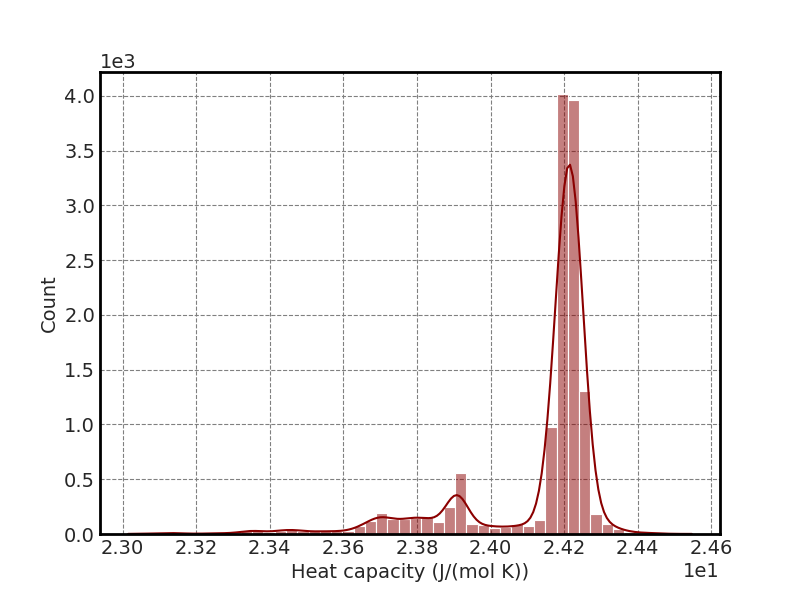}
        \caption{Heat Capacity}
        \label{fig:heat_cap_dist}
    \end{subfigure}
    \hfill
    \begin{subfigure}[b]{0.49\textwidth}
        \centering
        \includegraphics[width=\textwidth]{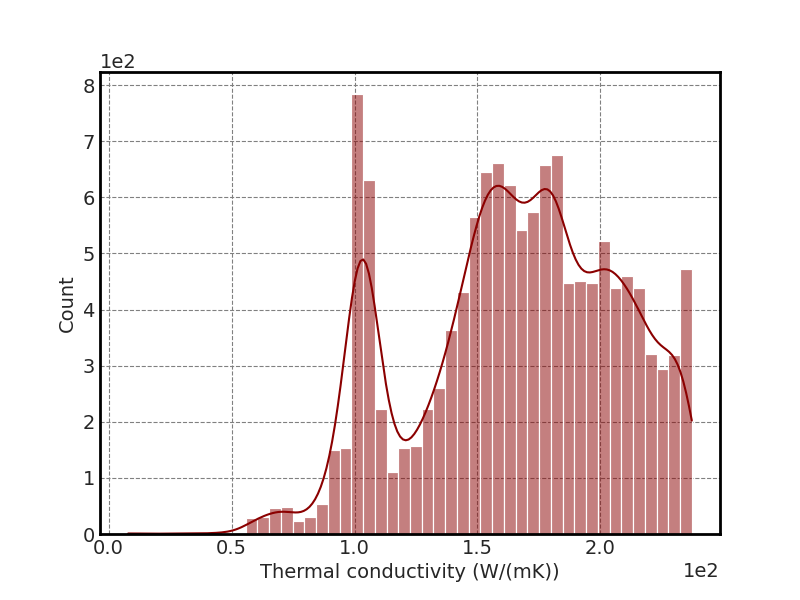}
        \caption{Thermal Conductivity}
        \label{fig:thermal_cond_dist}
    \end{subfigure}
    \caption{Target property distribution of the computational dataset. (a) Density, (b) Electrical Conductivity, (c) Heat Capacity, and (d) Thermal Conductivity}
    \label{fig:comp_data_hist}
\end{figure}

\begin{figure}
    \centering
    \captionsetup{justification=centering}
    \includegraphics[width=0.8\textwidth]{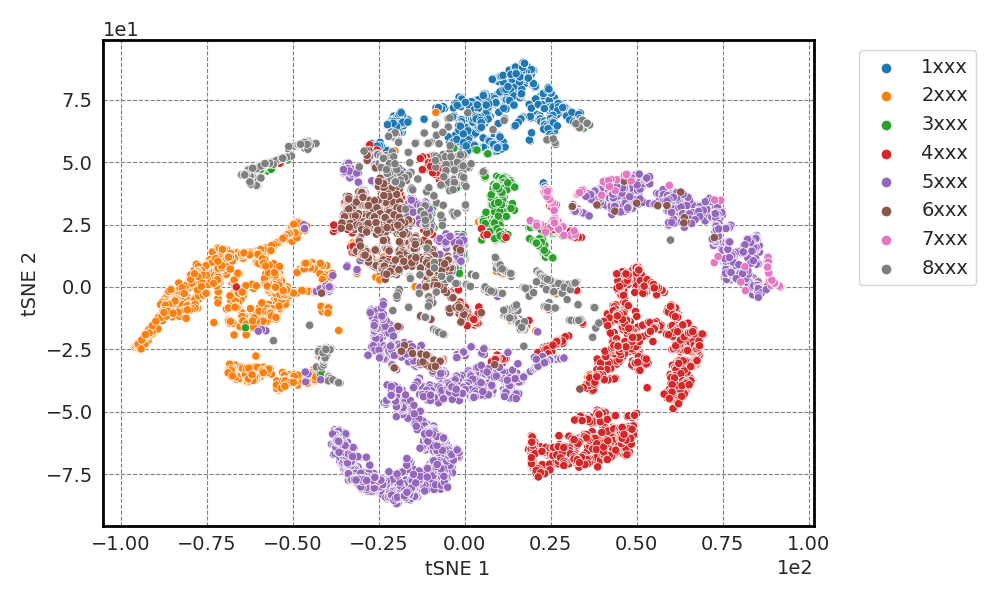}
    \caption{The t-Stochastic Neighbor Embedding (tSNE) illustrates the distribution of the computational data in the two-dimensional latent space. Different colours indicate the chemical composition of various series of aluminium alloy materials.}
    \label{fig:comp_tsne}
\end{figure}

\begin{figure}
    \centering
    \captionsetup{justification=centering}
    \begin{subfigure}[b]{0.49\textwidth}
        \centering
        \includegraphics[width=\textwidth]{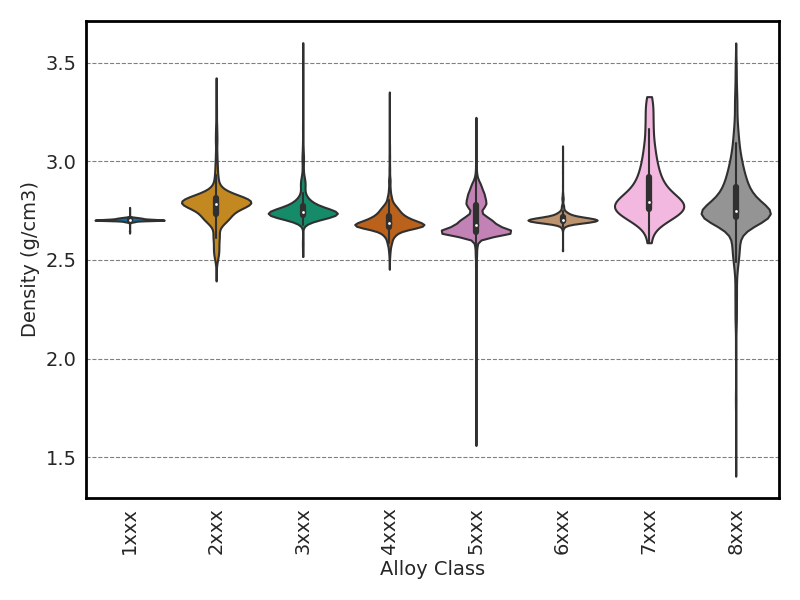}
        \caption{Density}
        \label{fig:density_violin}
    \end{subfigure}
    \hfill
    \begin{subfigure}[b]{0.49\textwidth}
        \centering
        \includegraphics[width=\textwidth]{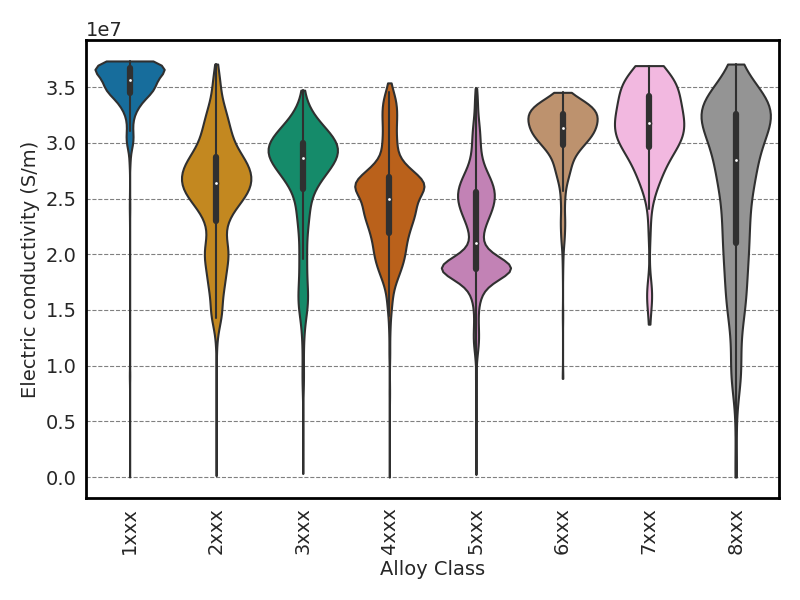}
        \caption{Electrical Conductivity}
        \label{fig:elec_cond_violin}
    \end{subfigure}
    \hfill
    \begin{subfigure}[b]{0.49\textwidth}
        \centering
        \includegraphics[width=\textwidth]{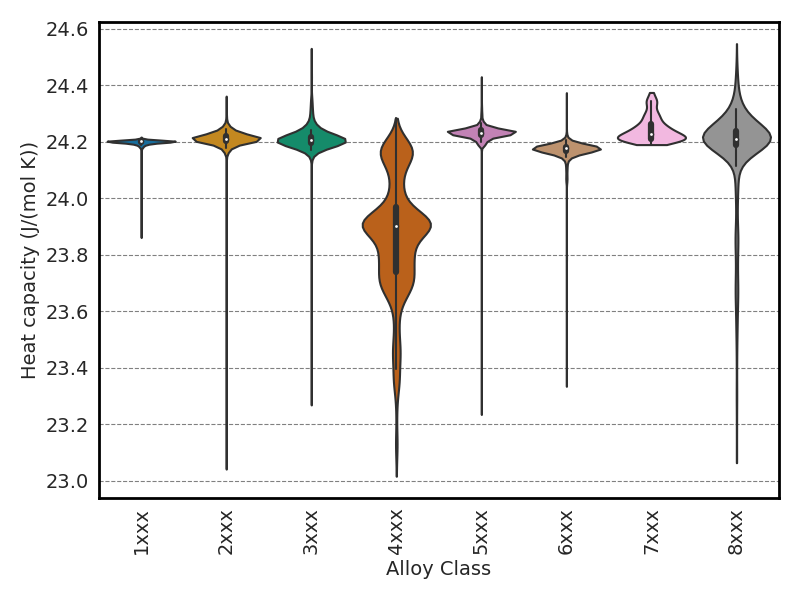}
        \caption{Heat Capacity}
        \label{fig:heat_cap_violin}
    \end{subfigure}
    \hfill
    \begin{subfigure}[b]{0.49\textwidth}
        \centering
        \includegraphics[width=\textwidth]{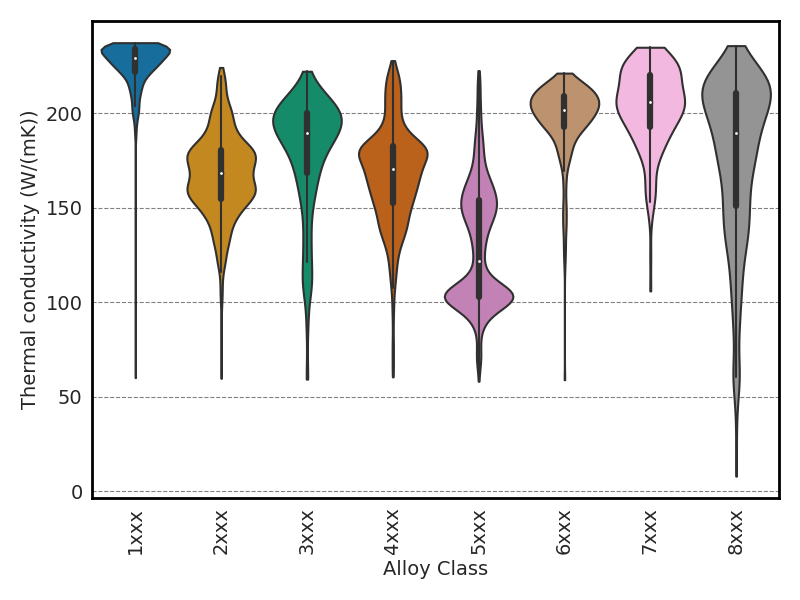}
        \caption{Thermal Conductivity}
        \label{fig:thermal_cond_violin}
    \end{subfigure}
    \caption{Material Property Distribution in the Computational Dataset. Various colours represent distinct classes of Aluminium alloy materials. The violin plot displays the distribution of (a) Density, (b) Electrical Conductivity, (c) Heat Capacity, and (d) Thermal Conductivity, categorized by the alloy series.}
    \label{fig:comp_data_violin}
\end{figure}

Supporting Figure \ref{fig:comp_data_hist} displays a histogram illustrating the
distribution of target properties in the computational data. In Supporting
Figure \ref{fig:comp_tsne}, we illustrate a 2D projection of the high-dimensional
composition space for aluminium alloys from computational data with the
assistance of the t-Stochastic Neighbor Embedding (tSNE) plot. Here, the aluminium
alloys are classified into distinct series based on their composition: the 1xxx
series primarily composed of at least 99\% aluminium; the 2xxx series featuring
copper as the principal alloying element; the 3xxx series characterised by
manganese as the primary alloying component; the 4xxx series, where silicon
serves as the primary alloying agent; the 5xxx series distinguished by magnesium
as the main alloying element; the 6xxx series comprising silicon and magnesium
as the primary alloying elements; the 7xxx series characterised by zinc as the
primary alloying element, alongside magnesium and copper; and finally, the 8xxx
series encompassing miscellaneous aluminium alloys not classified within the
preceding series. The colour of the points in Figure \ref{fig:comp_tsne}
indicates crucial alloy composition information for aluminium alloys derived
from computational data, where the colour corresponds to the alloy series
(1xxx is blue, 2xxx is dark yellow, 3xxx is green, 4xxx is orange, 5xxx is purple,
6xxx is brown, 7xxx is pink, and 8xxx is grey). Furthermore, in Supporting Figure
\ref{fig:comp_data_violin}, we present a violin plot
to demonstrate the distribution of computational material properties (e.g., density,
electrical conductivity, heat capacity, and thermal conductivity), categorised by
the alloy series.

\section*{Appendix B}

\begin{figure}
    \centering
    \captionsetup{justification=centering}
    \begin{subfigure}[b]{0.49\textwidth}
        \centering
        \includegraphics[width=\textwidth]{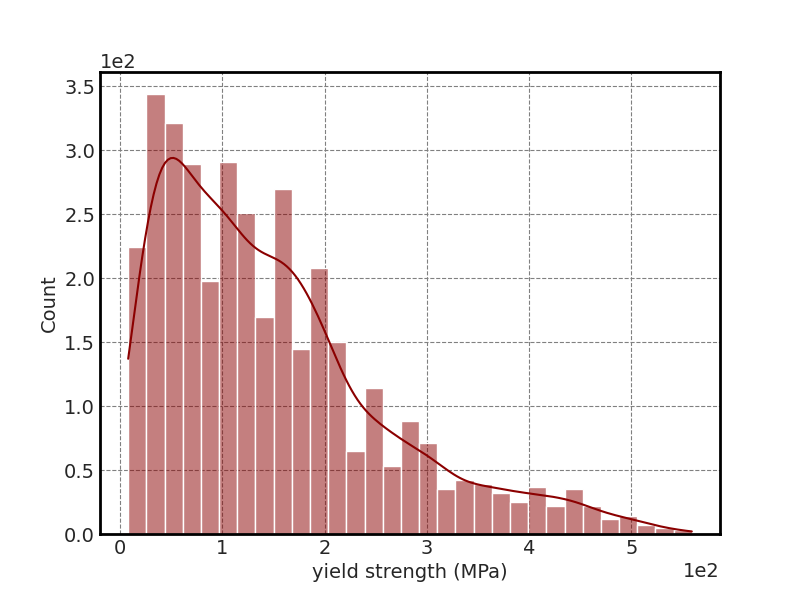}
        \caption{Yield Strength}
        \label{fig:yieldstrength_dist}
    \end{subfigure}
    \hfill
    \begin{subfigure}[b]{0.49\textwidth}
        \centering
        \includegraphics[width=\textwidth]{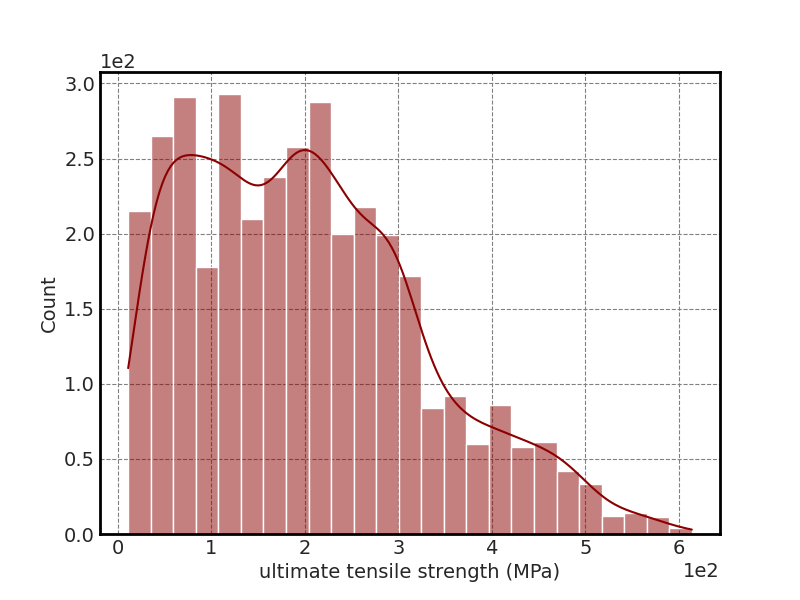}
        \caption{Ultimate Tensile Strength}
        \label{fig:ultimatetensilestrength_dist}
    \end{subfigure}
    \hfill
    \begin{subfigure}[b]{0.49\textwidth}
        \centering
        \includegraphics[width=\textwidth]{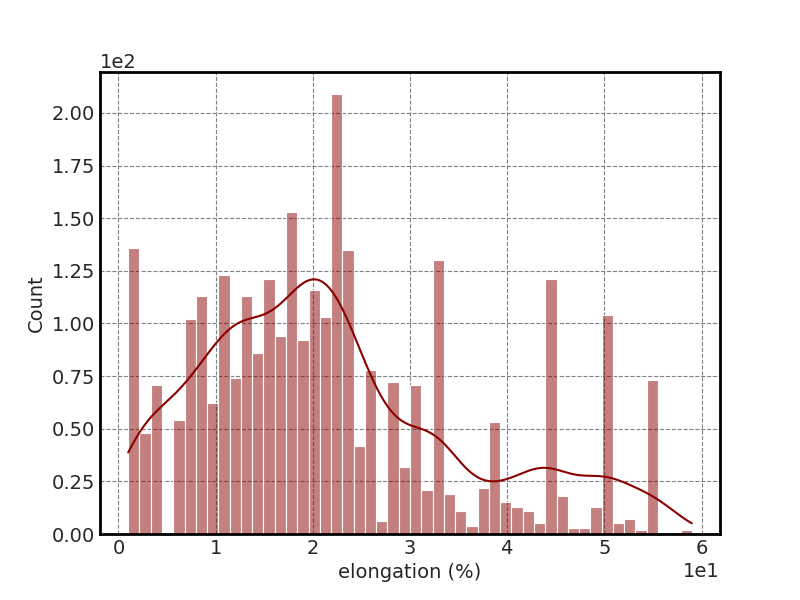}
        \caption{Elongation}
        \label{fig:elongation_dist}
    \end{subfigure}
    \caption{Target property distribution of the experimental dataset. (a) Yield Strength, (b) Ulitimate Tensile Strength, and (c) Elongation}
    \label{fig:exp_data_hist}
\end{figure}

\begin{figure}
    \centering
    \captionsetup{justification=centering}
    \includegraphics[width=0.8\textwidth]{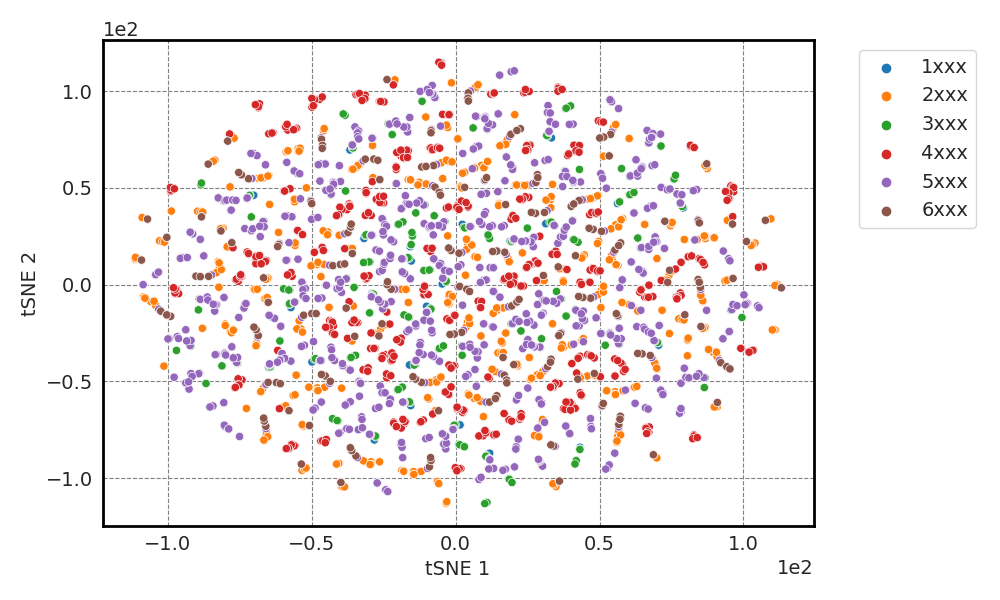}
    \caption{The t-Stochastic Neighbor Embedding (tSNE) illustrates the distribution of various data instances projected on a 2D plane from the experimental data. Different colours indicate the chemical composition of different series of aluminium alloy materials.}
    \label{fig:exp_tsne}
\end{figure}

\begin{figure}
    \centering
    \captionsetup{justification=centering}
    \begin{subfigure}[b]{0.49\textwidth}
        \centering
        \includegraphics[width=\textwidth]{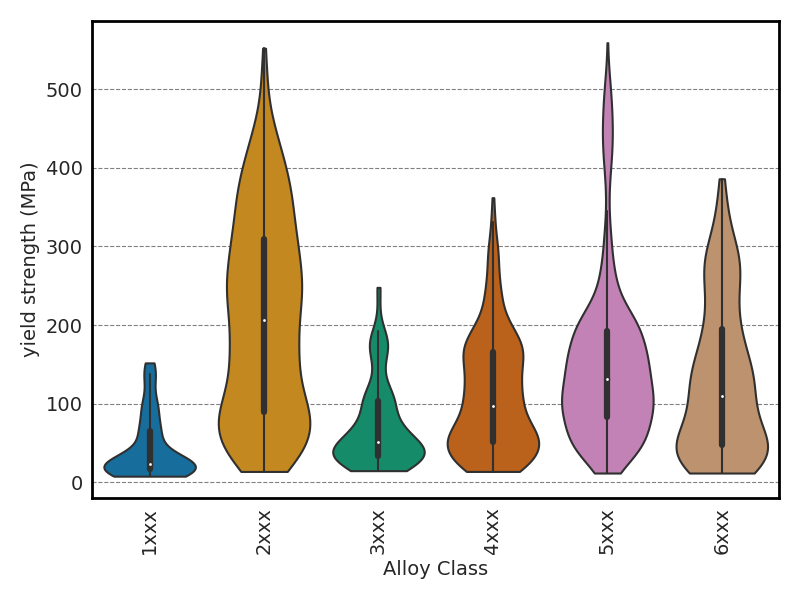}
        \caption{Yield Strength}
        \label{fig:yieldstrength_violin}
    \end{subfigure}
    \hfill
    \begin{subfigure}[b]{0.49\textwidth}
        \centering
        \includegraphics[width=\textwidth]{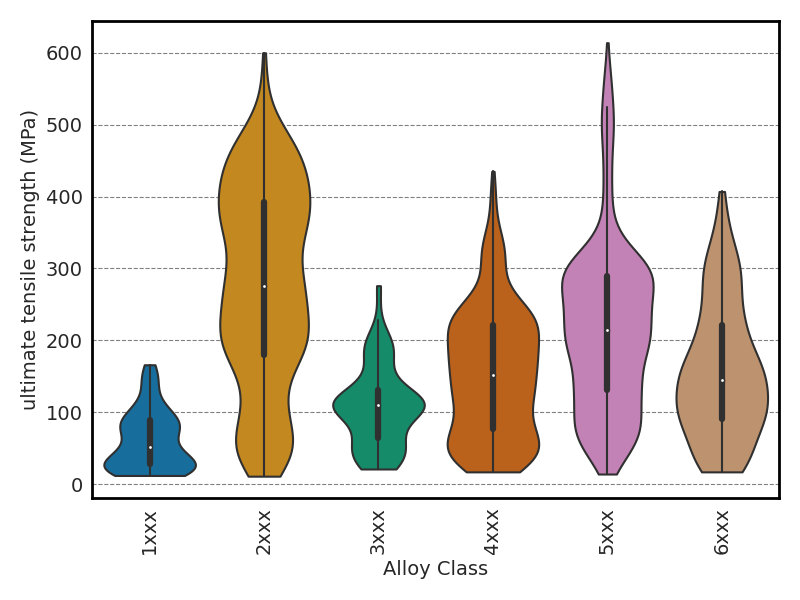}
        \caption{Ultimate Tensile Strength}
        \label{fig:ultimatetensilestrength_violin}
    \end{subfigure}
    \hfill
    \begin{subfigure}[b]{0.49\textwidth}
        \centering
        \includegraphics[width=\textwidth]{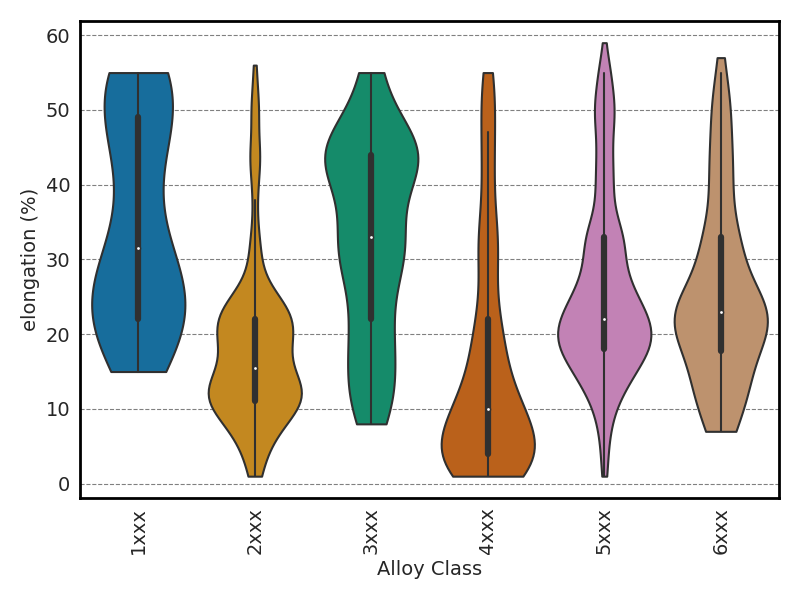}
        \caption{Elongation}
        \label{fig:elongation_violin}
    \end{subfigure}
    \caption{Distribution of Mechanical Properties in the Experimental Dataset: Different colours indicate distinct classes of aluminium alloy materials. The violin plot showcases the distribution of (a) Yield Strength, (b) Ultimate Tensile Strength, and (c) Elongation, classified by the alloy series.}
    \label{fig:exp_data_violin}
\end{figure}

Supporting Figure \ref{fig:exp_data_hist} shows a histogram illustrating the distribution
of these target properties in the experimental data. Supporting Figures \ref{fig:exp_tsne}
and \ref{fig:exp_data_violin}, respectively, depict a 2D projection of the
high-dimensional feature space for aluminium alloys from experimental data using
the tSNE plot and a violin plot illustrating the distribution of the experimental
material properties (e.g., yield stress, ultimate tensile stress, and elongation),
categorised by the alloy series.

